\documentclass[a4paper,11pt]{article}
\usepackage{jheppub}
\usepackage{amsmath,amssymb,slashed,braket}
\usepackage{graphicx}
\usepackage{lineno}
\usepackage{mathrsfs}

\title{\boldmath Dark-pion dark matter beyond leading order: unitarized chiral dynamics}

\author{Yuki Watanabe}
\affiliation{Kavli IPMU (WPI), UTIAS, University of Tokyo, Kashiwa, Chiba 277-8583, Japan}

\emailAdd{yuki.watanabe@ipmu.jp}

\abstract{
Dark pions are promising dark matter candidates, yet most analyses rely on leading-order (LO) chiral perturbation theory (ChPT). Motivated by the fact that, even for QCD $\pi\pi$ scattering, LO ChPT near threshold underestimates the isoscalar $s$-wave amplitude by an $\mathcal{O}(1)$ factor relative to high-precision dispersive results, we quantify how unitarization modifies the standard LO ChPT picture using the chiral unitary method—a nonperturbative resummation that implements the correct analytic structure with minimal input—and assess its impact on the phenomenology of dark-pion DM, taking SIMP and WIMP scenarios as canonical examples. We fix the subtraction constant to its natural estimate, interpreted as an effective cutoff at $\Lambda_\chi = 4\pi f_\pi$, so that the unitarized amplitudes depend only on $m_\pi$ and $f_\pi$. We show that, depending on the coupling $m_\pi/f_\pi$, the unitarized amplitudes develop resonance poles absent at LO, leading to sizable departures in $2\to2$ self-scattering, relevant for SIMP scenarios, and in annihilation including initial-state interaction effects, relevant for WIMP scenarios. These modifications, in turn, affect the viable parameter space. Although the subtraction constant is, from a model-building perspective, merely a parameter, a substantial deviation from its natural value would point to additional elementary resonances with the same quantum numbers.
}

\begin{document}
\maketitle
\flushbottom

\section{Introduction}

Echoes of the Big Bang bear witness to a nonluminous component—dark matter (DM)—yet its visage remains veiled\,\cite{Cirelli:2024ssz}. Since DM constitutes a significant portion of the matter in the universe, it is natural to consider that it may be structured through interactions analogous to those governing ordinary matter, such as quantum chromodynamics (QCD). A plausible realization of this idea is a confining, QCD-like dark sector whose low-energy degrees of freedom parallel hadronic dynamics and whose composite states, “dark hadrons”, constitute DM\,\cite{Kribs:2016cew,Cline:2021itd}. In close analogy with QCD, it is then natural to focus on scenarios in which the lowest-lying excitations of the dark sector—the counterparts of QCD pions—play the role of DM, commonly referred to as “dark pions”\,\cite{Hochberg:2014kqa,Hochberg:2015vrg,Hochberg:2018rjs,Choi:2016hid,Lee:2015gsa,Bhattacharya:2013kma,Harigaya:2016rwr,Cheng:2021kjg,Abe:2024mwa,Arina:2019tib}.

Once such a strongly coupled dark sector is assumed, however, directly analyzing its UV Lagrangian faces the same difficulties as QCD: reliably accessing the strongly interacting IR regime, where straightforward perturbation theory fails, requires genuinely nonperturbative tools. In practice, as in ordinary hadron physics, a fully controlled treatment of the UV-complete theory would call for methods such as lattice simulations, while an analytic derivation of the dark-hadron spectrum by first-principles methods remains, at present, beyond our capabilities. For the purpose of describing the phenomenology of composite DM, it is therefore economical to employ an effective field theory (EFT) formulated directly in terms of the composite degrees of freedom that are expected to be relevant at the scales of interest\,\cite{Manohar:1996cq,Kaplan:2005es,Meissner:2022cbi}.

Even after one passes to such an EFT description, the analysis can remain highly nontrivial in the presence of strong dynamics. A textbook example is nonrelativistic nucleon–nucleon scattering\,\cite{Kaplan:1998tg,Kaplan:1998we,Bedaque:2002mn,Chen:1999tn,vanKolck:1999mw}. At sufficiently low momenta, the de Broglie wavelength of the nucleons exceeds the range of the nuclear force, so that the detailed structure of the potential is not resolved and the interaction can be approximated by contact terms. One might therefore expect that a Lagrangian with contact interactions and a perturbative expansion in those couplings would suffice. Yet empirically, the $s$-wave scattering lengths are much larger than the interaction range and are controlled by shallow bound or virtual states near threshold. Under these conditions, a finite-order expansion in the contact interactions is simply inapplicable; the contact interactions must be resummed nonperturbatively to obtain unitary amplitudes. In Wilsonian renormalization-group terms, low-energy nucleon physics is governed not by the Gaussian fixed point of ``natural'' scattering lengths but by a nontrivial fixed point associated with the unitary limit\,\cite{Birse:1998dk,Barford:2002je}.

These considerations suggest that IR EFTs generated by strong dynamics may generically require nonperturbative resummations of scattering amplitudes, rather than naive truncations of a perturbative series. This, in turn, motivates asking whether similar issues arise in composite DM frameworks. In this work we explore this question in the specific setting of dark-pion DM, in which the DM is identified with the lowest-lying excitations of a QCD-like strong sector. By “dark pions” we mean the analogues of QCD pions in a dark sector: (pseudo-)Nambu–Goldstone bosons associated with the spontaneous breaking of an approximate global chiral symmetry of the UV strong dynamics by dark-quark condensation in the IR. The structure of their interactions is dictated by this symmetry and its breaking pattern, and is encoded in chiral perturbation theory (ChPT), which provides the corresponding low-energy EFT.

At the model-building level, quantitative assessments of dark-pion DM are most often carried out at leading order (LO) in ChPT\,\cite{Weinberg:1968de,Weinberg:1978kz,Gasser:1983yg,Gasser:1984gg,Pich:1995bw,Scherer:2012xha}, using tree-level amplitudes for dark-pion self-scattering and annihilation. A natural question is whether LO ChPT is a sufficiently reliable starting point for the observables of interest. Experience with real QCD pions suggests caution. For $\pi\pi$ scattering, LO ChPT does not reproduce experimental data with high precision even near threshold: in the isoscalar $s$-wave channel, the scattering length extracted from experiment exceeds the LO prediction by an $\mathcal{O}(1)$ factor, corresponding to roughly a factor-of-two discrepancy in the near-threshold cross section.

One obvious response is to compute ChPT to higher orders\,\cite{Bijnens:2011fm}. Indeed, in some channels the next-to-leading and next-to-next-to-leading corrections can be sizable. However, going to higher orders introduces an ever-growing number of low-energy constants (LECs), and because ChPT is formulated as a momentum expansion, the convergence of the perturbative series rapidly deteriorates as the momentum increases, so that even high-order calculations, with many additional LECs to be fixed, retain only a relatively narrow kinematic range of validity.
Moreover, as long as only pions are kept as explicit degrees of freedom and the expansion is truncated at finite order, the spectrum of light meson resonances seen experimentally—most notably the broad $\sigma$ in the isoscalar $s$-wave channel and the $\rho$ in the isovector $p$-wave channel—is not reproduced, even though the $\sigma$ significantly affects near-threshold scattering lengths and both resonances leave a pronounced imprint on the observed phase shifts\,\cite{ParticleDataGroup:2024cfk}.
Analogous considerations apply in the dark sector. One may attempt to estimate or fit higher-order LECs\,\cite{Kolesova:2025ghl,Hansen:2015yaa}, appeal to lattice calculations\,\cite{Kulkarni:2022bvh,Zierler:2022uez,Dengler:2024maq}, or employ holographic models\,\cite{Alfano:2025non}. However, because the symmetry-breaking pattern and the underlying strong dynamics of the dark sector are a priori unknown, such approaches inevitably become model-specific and do not provide a uniform, systematic handle across different classes of theories.

For these reasons, the necessity of employing unitarization methods—nonperturbative resummations of ChPT scattering amplitudes that restore exact two-body unitarity while preserving the chiral low-energy expansion—has long been emphasized in hadron physics\,\cite{Oller:2020guq}. A standard example is the inverse amplitude method (IAM)\,\cite{Dobado:1996ps,GomezNicola:2007qj}, which uses one-loop (NLO) ChPT amplitudes as input to construct unitarized partial waves that dynamically generate resonances. However, IAM requires ChPT amplitudes beyond LO; in dark-pion ChPT, where higher-order calculations quickly become technically demanding and the corresponding LECs are often poorly constrained, IAM is not necessarily the most economical choice for a minimal and broadly applicable framework.

In this work we instead adopt, as a minimal step beyond LO, the so-called chiral unitary method\,\cite{Oller:1998hw,Oller:1998zr,Oller:2000ma,Oller:2019opk} and investigate how unitarization affects DM scenarios realized by dark pions. The chiral unitary approach starts from the LO ChPT amplitude and nonperturbatively completes it in such a way that two-body unitarity and analyticity are satisfied. In practice, this is achieved via a dispersive construction characterized by a single subtraction constant, which plays the role of a renormalization parameter. Although this parameter is in principle model dependent, one can employ a natural-value estimate to relate it to an effective cutoff of the dark-pion EFT. In this way, the nonperturbative completion preserves the minimal structure of the LO theory while dynamically generating resonance poles and endowing the scattering amplitude with the correct analytic structure.

We apply this chiral-unitary method to two representative realizations of dark-pion DM, namely SIMP\,\cite{Hochberg:2014dra} and WIMP\,\cite{Arcadi:2017kky} scenarios, and examine its implications for DM phenomenology. In the SIMP case, the relic abundance is set by $3\to 2$ freeze-out, so that the required coupling naturally induces large DM self-interactions, which may help address small-scale structure problems\,\cite{Spergel:1999mh,Tulin:2017ara}. Our analysis focuses on how much the dark-pion self-scattering cross sections, as predicted at LO, can be modified by unitarization. In the WIMP case, the impact of unitarization manifests itself as rescattering via dark-pion self-interactions in the annihilation process. In hadron physics, such effects are known as initial-state interactions\,\cite{Oller:2019opk, Oller:2020guq} and can be viewed as the analogue of Sommerfeld enhancement in scenarios with a light mediator\,\cite{Hisano:2002fk,Hisano:2003ec,Hisano:2004ds,Arkani-Hamed:2008hhe}.

The paper is organized as follows. Sec.~\ref{sec:chiral perturbation} briefly reviews ChPT and partial-wave analysis as applied to QCD pion scattering. Sec.~\ref{sec:chiral dynamics} introduces the chiral unitary method, compares it with tree-level ChPT for QCD pions, and tracks how the amplitude and resonance poles evolve as $m_\pi/f_\pi$ are varied in the dark sector. Sec.~\ref{sec:Implication} presents the implications for dark-pion DM in SIMP and WIMP scenarios, emphasizing self-scattering and annihilation amplitudes relevant for relic abundance and phenomenology. We conclude in Sec.~\ref{sec:conclusion}.    
\section{Chiral perturbation theory}
\label{sec:chiral perturbation}

As groundwork for the dark-pion system, this section briefly summarizes chiral perturbation theory and partial-wave analysis for describing pion scattering, and presents the leading-order perturbative predictions. For further details, see\,\cite{Gasser:1983yg,Gasser:1984gg,Pich:1995bw,Scherer:2012xha}.

\subsection{Chiral Lagrangian}

A dark pion is a (pseudo-)Nambu–Goldstone (NG) boson that arises when an (approximate) global symmetry $G$ of the underlying QCD-like dark sector is spontaneously broken to a subgroup $H$. In general, the structure of the low-energy effective theory for NG bosons is determined by the symmetry breaking pattern $G/H$ through the CCWZ construction\,\cite{Coleman:1969sm,Callan:1969sn}. A canonical example is provided by the lowest excitation of QCD: the pions arise from the spontaneous breaking of the approximate chiral symmetry of the light $u$ and $d$ quarks, $SU(2)_L \times SU(2)_R \to SU(2)_V$, induced by the quark condensate $\braket{\bar{q}q}$. Writing the interpolating field of $\pi$ as a coset representative with the NG coordinates, $U(x)=\exp(2i\pi^a(x)T^a/f_\pi)$ with $f_\pi$ being the pion decay constant and $T^a$ being the broken generators normalized by $\mathrm{Tr}(T^a T^b)=\delta^{ab}/2$, the leading-order effective Lagrangian of pions is given by
\begin{align}
    \label{eq:chiral lagrangian}
    \mathcal{L}^{(2)} = \frac{f_\pi^2}{4} \mathrm{Tr} \big( \partial_\mu U \, \partial^\mu U^\dagger \big)
    + \frac{f_\pi^2}{4}\mathrm{Tr}\big(\chi U^\dagger+\chi^\dagger U\big).
\end{align}
Here, $\chi$ is a spurion field that tracks the explicit breaking of $G$; once it is assigned a nonzero background value, the would-be Nambu–Goldstone fields $\pi^a(x)$ acquire masses. Indeed, one can write $\chi=2BM$, 
with the quark mass matrix $M=\mathrm{diag}(m_u,m_d)$, and matching to QCD yields the GMOR relation 
$m_\pi^2 f_\pi^2 = 2 B (m_u+m_d) f_\pi^2 = - (m_u+m_d) \braket{\bar{q}q}$ for the pion mass $m_\pi$\,\cite{Gell-Mann:1968hlm,Gasser:1983yg}. In the following, we simply take $\chi = m_\pi^2 \mathbf{1}$.

Higher-order terms in the Lagrangian are also systematically constructed by the CCWZ construction and organized with the appropriate power counting scheme, the chiral counting, and the resulting perturbative series including the scattering amplitudes are renormalized order by order. This framework for the effective field theory of $\pi$ is known as the chiral perturbation theory (ChPT)\,\cite{Weinberg:1968de,Weinberg:1978kz,Gasser:1983yg,Gasser:1984gg,Pich:1995bw,Scherer:2012xha}. Dark pion models can also be described within ChPT once the symmetry-breaking pattern $G/H$ of the dark sector is specified. Representative examples include $SU(N_f)_L\times SU(N_f)_R \to SU(N_f)_V$, which is the QCD-like case with $N_f$ dark quarks in complex representations; $SU(2N_f)\to SO(2N_f)$, arising for fermions in real representations; and $SU(2N_f)\to Sp(2N_f)$, characteristic of pseudoreal representations, giving the cosets $SU(N_f)_L\times SU(N_f)_R/SU(N_f)_V$, $SU(2N_f)/SO(2N_f)$, and $SU(2N_f)/Sp(2N_f)$, respectively. The scattering amplitudes and the group-theoretical factors needed to compute them can be found in Ref.~\cite{Hochberg:2014kqa,Kamada:2022zwb} for various symmetry-breaking patterns.

\subsection{Partial wave analysis}

In practice, it is convenient to decompose the scattering amplitude into partial waves, namely, components of definite angular momentum. For a $2\to2$ process, the invariant amplitude $\mathcal{M}(s,t,u)$ depends on the Mandelstam variables $s, t$ and $u$, but by kinematics it can be rewritten as a function of the collision energy $\sqrt{s}$ and the scattering angle $\theta$ with respect to the collision axis. We then define the partial-wave expansion of the scattering amplitude with respect to orbital angular momentum $\ell$ by
\begin{align}
    \label{eq:partial wave}
    \mathcal{M}(s,\cos\theta ) &= 2\sum^\infty_{\ell=0}(2\ell+1)P_\ell(\cos \theta) T_\ell(s),
\end{align}
where $P_\ell(x)$ are the Legendre polynomials and $T_\ell(s)$ is the $\ell$-th partial wave amplitude. The overall factor of 2 accounts for identical particles. In particular, with the partial-wave decomposition, the self-scattering cross section $\sigma_{\rm el}(s)$ splits into contributions from each partial wave as
\begin{align}
    \label{eq:elastic cross section partial wave}
    \sigma_{\rm el}(s) = \sum_{\ell=0}^\infty \sigma_\ell(s), \quad 
    \sigma_\ell(s) &= \frac{2\ell+1}{8\pi s} |T_\ell(s)|^2.
\end{align}

An advantage of the partial-wave decomposition is that the nontrivial unitarity constraint on the $S$-matrix, namely, the optical theorem for the scattering amplitude, takes a simpler form than in the momentum-space representation. The optical theorem implies, for the forward scattering amplitude of two particles with equal mass $m$,
\begin{align}
    {\rm Im}\,\mathcal{M}_{2\to2}\,(\theta=0) = 2\sqrt{s}\,|\vec{p}|\, \sigma^{\rm tot}(s),
\end{align}
where $|\vec{p}| = \sqrt{s-4m^2}/2$ is the magnitude of the momentum in the center-of-mass frame\,\footnote{For incoming particles with unequal masses $m_1$ and $m_2$, $|\vec{p}| = \lambda^{1/2}(s, m_1^2, m_2^2) / (2\sqrt{s})$ with $\lambda(x, y, z) = x^2 + y^2 + z^2 - 2xy - 2yz - 2zx$ being the K\"{a}llén triangle function.}  and $\sigma^{\rm tot}(s)$ is the total cross section of the two-particle initial state. Combining this with the partial-wave expansion in Eq.~(\ref{eq:partial wave}), the relation simplifies, by virtue of the orthogonality of the Legendre polynomials $P_\ell(x)$, to an algebraic condition for each angular momentum $\ell$. While Eq.~(\ref{eq:optical theorem}) assumes a fixed initial state, the optical theorem can be extended to the multichannel case by allowing the initial state to be coupled two-body channels as well.
The corresponding partial-wave unitarity condition then reads
\begin{align} 
    \label{eq:optical theorem}
    {\rm Im}\, T_{\ell}(s) = T_\ell^\dagger(s)\rho(s)T_\ell(s), \quad \rho(s) = {\rm diag}\,(\rho_1(s),\rho_2(s),\cdots).
\end{align}
Here the $T$-matrix element $T_{\ell,ij}(s)$ denotes the partial-wave scattering amplitude from channel $i$ to channel $j$. The phase-space factor $\rho_i(s)$ is given by $\rho_i(s)=|\vec{p}_i|/8\pi\sqrt{s}$, where $|\vec{p}_i|$ is the center-of-mass momentum of channel $i$. This unitarity relation applies only to channels that are kinematically open at the given $\sqrt{s}$.

Note that dark pions transform in a linear representation of the unbroken subgroup $H$, and the Lorentz-invariant scattering amplitude $\mathcal{M}$ carries the corresponding flavor indices. Upon decomposing the two-particle state into irreducible representations $I$ of $H$ and expanding in partial waves, the $S$-matrix becomes fully diagonal in that basis, and the optical theorem in Eq.~(\ref{eq:optical theorem}) is imposed for the $(\ell,I)$ scattering amplitude $T_\ell^I(s)$ channel by channel.

The optical theorem in Eq.~(\ref{eq:optical theorem}) for the elastic partial-wave amplitude $T_\ell^I(s)$ can be written in terms of the phase shift $\delta_\ell^I(s)$ as
\begin{align}
    \label{eq:amplitude and phase shift}
    T^I_\ell(s) &= 8\pi\sqrt{s}\frac{|\vec{p}|^{2\ell}}{|\vec{p}|^{2\ell+1}\cot \delta_\ell^I(s)-i|\vec{p}|^{2\ell+1}}.
\end{align}
The factor $|\vec{p}|^{2\ell}$ in the numerator governs the threshold behavior of the $\ell$-th partial-wave amplitude, while the imaginary part in the denominator ensures elastic unitarity. Because $T_\ell^I(s) \propto |\vec{p}|^{2\ell}$ as $|\vec{p}|\to 0$, higher partial waves are power-suppressed near threshold, so that only the lowest few waves contribute appreciably at low energies—one of the practical advantages of the partial-wave expansion. The nontrivial part of the interaction is thus contained in the remaining term $|\vec{p}|^{2\ell+1}\cot \delta^I_\ell(s)$, which admits the following effective range expansion near the threshold\,\cite{Bethe:1949yr}:
\begin{align}
    \label{eq:ERE}
    |\vec{p}|^{2\ell+1}\cot \delta_\ell^I(s) = -\frac{1}{A_\ell^I}+\frac{1}{2}r^I_\ell |\vec{p}|^2 + \mathcal{O}(|\vec{p}|^4),
\end{align}
where $A^I_\ell$ and $r^I_\ell$ are the scattering length and the effective range of the $(\ell,I)$ partial wave, respectively.\,\footnote{As discussed later, since the $\pi \pi$ amplitude contains a zero, i.e., a CDD pole, this expansion is not necessarily accurate when the pole’s effect is significant\,\cite{Oller:2019opk}.} In particular, the scattering length $A_\ell^I$ provides a characterization of low-energy scattering near the threshold in the $(\ell,I)$ channel. 

\subsection{Leading order predictions}

Let us consider elastic $\pi \pi$ scattering in QCD within ChPT, where the Lagrangian is given by Eq.~(\ref{eq:chiral lagrangian}).
For $m_\pi \simeq 138\,\mathrm{MeV}$ and $f_\pi \simeq 92.3\,\mathrm{MeV}$, the coupling constant of ChPT, $m_\pi/f_\pi \simeq 1.50$, is much smaller than the naive perturbative bound $4\pi$. Together with the derivative interactions of pions, this might suggest that the tree-level or the leading-order (LO) approximation should be reliable—at least near threshold, where the pions are nonrelativistic. Nevertheless, as we shall see, this expectation fails even in that regime.

The invariant amplitude for $\pi^a\pi^b \to \pi^c\pi^d$ scattering, with $a,b,c,d = 1,2,3$ being the flavor indices, where the pions transform in the adjoint representation of the unbroken subgroup $SU(2)_V$, i.e., an isotriplet, has the following structure:
\begin{align}
    \mathcal{M}^{ab;cd}(s,t,u) = 
    A(s,t,u)\delta^{ab}\delta^{cd} 
    +
    A(t,s,u)\delta^{ac}\delta^{bd}+A(u,t,s)\delta^{ad}\delta^{bc}.
\end{align}
The leading-order chiral Lagrangian in Eq.~(\ref{eq:chiral lagrangian}) yields the tree-level result $A(s,t,u) = (s - m_\pi^2)/{f_\pi^2}$. As already noted, the invariant amplitude can be decomposed into the $(\ell,I)$ basis $T^I_\ell(s)$. By projecting the two-pion state onto total isospin $I=0,1,2$ and then performing the partial-wave expansion Eq.~(\ref{eq:partial wave}) in the total angular momentum $\ell$, we obtain
\begin{align}
    \label{eq:LO amplitudes}
    T_0^0(s) =\frac{s-m_\pi^2/2}{f_\pi^2}, \quad T_1^1(s) =\frac{s-4m_\pi^2}{6f_\pi^2}, \quad
    T_0^2(s) =-\frac{s-2m_\pi^2}{2f_\pi^2}.
\end{align}
The $p$-wave amplitude $T_1^1(s)$ is kinematically suppressed and vanishes at threshold, behaving as $|\vec{p}|^2$, while the $s$-wave amplitudes $T_0^0(s)$ and $T_0^2(s)$ likewise possess a zero, known as the Adler zero\,\cite{Adler:1964um,Weinberg:1966kf}. These zeros reflect the pion’s nature as a NG boson.

As a near-threshold observable within the domain of validity of ChPT, one can consider the scattering length in units of the pion mass, $a_\ell^I = -m_\pi A^I_\ell$. Combining Eq.~(\ref{eq:amplitude and phase shift}) and Eq.~(\ref{eq:ERE}) with the LO amplitudes in Eq.~(\ref{eq:LO amplitudes}) yields the famous Weinberg prediction\,\cite{Weinberg:1966kf} for the $s$-wave scattering lengths as:
\begin{align}
    \label{eq:Weinberg}
    a_0^{0} = \frac{7 m_\pi^2}{32\pi f_\pi^2}\simeq 0.16, 
    \quad
    a_0^{2} = -\frac{m_\pi^2}{16\pi f_\pi^2}\simeq -0.045.
\end{align}

On the other hand, high-precision determinations of $s$-wave scattering lengths, based on experimental data together with the state-of-the-art two-loop ChPT and dispersion relations (Roy equations\,\cite{Ananthanarayan:2000ht}), give\,\cite{Colangelo:2000jc}
\begin{align}
    \label{eq:scattering length Roy}
    a_0^{0} = 0.220 \pm 0.005, 
    \quad
    a_0^{2} = -0.0444 \pm 0.0010.
\end{align}
Comparing these with the LO results in Eq.~(\ref{eq:Weinberg}), one finds that the $I=2$ channel is already well described at LO, while the $I=0$ channel shows a discrepancy even at the threshold $s=4m_\pi^2$. The difference corresponds to a factor of about 1.9 in the cross section, which is phenomenologically large.  The origin of this difference is well understood. Because the $I=0$, $L=0$ channel is dominated by the broad isoscalar $\sigma$ resonance or the $f_0(500)$, whose influence extends over a wide energy range, including close to threshold, a simple perturbative expansion cannot capture its dynamics: tree-level ChPT underestimates its effect. While higher-order corrections are essential to predict the scattering length with reasonable accuracy, they remain insufficient to reproduce the full resonance structure. Moreover, in the $I=1$ channel the $\rho$ meson exists in the hadron spectrum, and it cannot be reproduced within simple ChPT. Consequently, one must go beyond a straightforward perturbative expansion and employ nonperturbative tools that resum higher-order corrections to the amplitudes and can incorporate resonance effects.

\section{Chiral dynamics}
\label{sec:chiral dynamics}

As noted above, LO ChPT, and even higher-order ChPT treated purely perturbatively, fails to fully match experimental data, particularly in the $I=0$ channel. The mismatch originates from the presence of the broad dynamical resonance, which affects the amplitude over a wide energy range and cannot be captured within a simple perturbative expansion.

To overcome these difficulties and to make ChPT quantitatively predictive over a broader kinematic range, it has long been emphasized that the perturbative series must be supplemented by some nonperturbative, unitarization method\,\cite{Oller:2020guq}: Since ChPT is a nonrenormalizable effective field theory, the number of operators and low-energy constants (LECs) proliferates rapidly with increasing chiral order, making systematic calculations impractical; even when higher orders are computed, the derivative expansion converges only within a narrow kinematic window and deteriorates as $\sqrt{s}$ grows; and resonance poles cannot be generated by any finite truncation. A controlled, nonperturbative resummation of ChPT, i.e., chiral dynamics, that respects the general properties of the scattering amplitude, namely, unitarity and analyticity, is therefore required.

These considerations apply with at least comparable severity to dark-pion models. The variety of possible patterns of symmetry breaking, comparable to or even more complex and diverse than those in QCD, inflates the number of required LECs and renders loop calculations substantially more demanding.
Phenomenologically, a relatively strong coupling $m_\pi/f_\pi$, compared to that of QCD pions, is often required to satisfy the relic abundance condition, which degrades the convergence of the perturbative series, and dynamically generated resonances may modify self-scattering and annihilation precisely near threshold, the key arena for DM physics. Consequently, LO-only treatments are unreliable, and a unitarized, analyticity-preserving analysis should be regarded as the standard starting point for dark-pion phenomenology.

Although the unitarization method is not unique, here we propose the chiral unitary method as a minimal extension of ChPT\,\cite{Oller:1998hw,Oller:1998zr,Oller:2000ma,Oller:2019opk}: it improves the applicability of ChPT while using only the LO amplitude together with a single subtraction/cutoff parameter, and it ensures both unitarity and analyticity.

\subsection{Chiral unitary method}

We introduce the chiral unitary method to improve the perturbative scattering amplitudes and make ChPT more predictive. Our treatment is kept minimal, and a more detailed discussion can be found in Refs.~\cite{Oller:1998hw,Oller:1998zr,Oller:2000ma,Oller:2019opk}.

\begin{figure}[t] 
    \centering
    \includegraphics[width=0.7\linewidth]{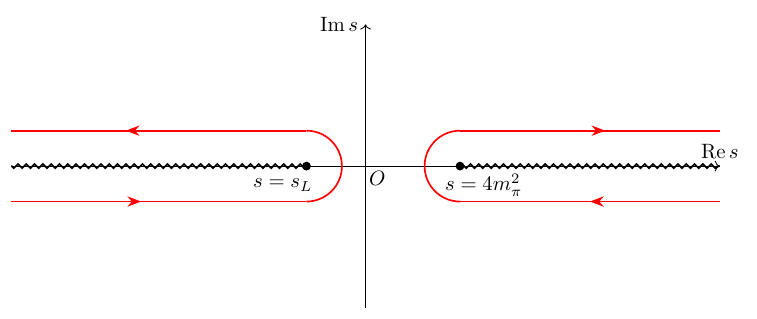}
    \caption{\small \sl  
    Analytic structure of the inverse partial-wave amplitude $T_\ell^{-1}(s)$. $s$-channel unitarity implies a right-hand or unitarity cut starting at threshold, while crossing symmetry implies a left-hand or dynamical cut, which is formally taken to start at $s=s_L$. The red lines indicate the integration contours used in the dispersion relation.
    }
    \label{fig:analytic structure of T inverse}
\end{figure}

We discuss the chiral unitary method for the elastic scattering of pions. An important property of the scattering amplitude, implied by causality, is that it is an analytic function of the complexified kinematic variables, whose singularity structure is constrained by unitarity. Then the starting point of the chiral unitary method is to rewrite the optical theorem in Eq.~(\ref{eq:optical theorem}) for the elastic channel, ${\rm Im}\,T_\ell(s) = \rho(s) |T_\ell(s)|^2$, as
\begin{align}
    \label{eq:optical inverse}
    {\rm Im}\, T_\ell^{-1}(s) = -\rho(s),
\end{align}
where $\rho(s) =\sqrt{1-4m_\pi^2/s}/16\pi$. This implies that the inverse scattering amplitude $T^{-1}_\ell(s)$, regarded as an analytic function of the Mandelstam variable $s$, has a unitarity cut beginning at the threshold $s_{\rm th} =  4m_\pi^2$. On the other hand, by crossing symmetry, the unitarity cuts in the $t$- and $u$-channels appear in the $s$-plane as the left-hand (dynamical) cut in the unphysical region. Accordingly, the analytic structure of $T_\ell^{-1}(s)$ in the complex $s$-plane is as shown in Fig.~\ref{fig:analytic structure of T inverse}. Here, we formally take the branch point of the left-hand cut to be at $s=s_L$, whose explicit location is not relevant for the following discussion.\,\footnote{For completeness, we note that in equal-mass $\pi\pi$ scattering the left-hand branch point is located at $s_L = 0$. This branch point originates from the crossed $t$- and $u$-channel thresholds at $t,u \ge 4m_\pi^2$. Using $
t=-(s-4m_\pi^2)(1-\cos\theta)/2$, one finds that the condition $t = 4m_\pi^2$ is first satisfied within the integration range $\cos\theta \in [-1,1]$ at $s = 0$. The same conclusion follows from the $u$-channel threshold, which yields an identical branch point $s_L=0$.} Physical values of the amplitude $T_\ell(s)$ are obtained as boundary values on the physical sheet, $T_\ell(s+i0^+)$; the branch cut of the square root in $\rho(s)$ is taken along the real axis. Hereafter, we always evaluate the physical amplitude with $s$ understood to carry a positive imaginary part, but we do not indicate this explicitly. 

Taking into account this analytic structure of $T_\ell^{-1}(s)$ and applying Cauchy’s integral formula on a circle around the origin, which can be deformed to infinity so that the arc contribution vanishes and only the branch–cut integrals remain (red contour in Fig.~\ref{fig:analytic structure of T inverse}), one can write the following dispersion relation:
\begin{align}
    \label{eq:chiral unitary}
    T^{-1}_{\ell}(s) = V_\ell^{-1}(s) + G(s).
\end{align}
Here, $V_\ell^{-1}(s)$ represents the dispersion integral along the left-hand cut, while $G(s)$ corresponds to the one along the right-hand cut, whose expression is given in Eq.~(\ref{eq:loop function}). Note that if the amplitude $T_\ell(s)$ vanishes at some value of $s$, its inverse $T_\ell^{-1}(s)$ becomes ill-defined. Such zeros, or poles of the inverse amplitude, are known as Castillejo-Dalitz-Dyson (CDD) poles\,\cite{Castillejo:1955ed}, and in our notation the contributions of these poles are also incorporated into $V_\ell(s)$. If the effect of the left-hand cut is sufficiently weak, it can be approximated by a polynomial in $s$; accordingly $V_\ell^{-1}(s)$ can be written as
\begin{align}
    V^{-1}_\ell(s) \simeq \sum_{n=0}^{N_L}a_ns^n + \sum_{i} \frac{\gamma_i}{s-s_i},
\end{align}
where the first term on the right-hand side represents the left-hand–cut contribution as a polynomial up to degree $N_L$, while the second term accounts for the CDD poles. In fact, due to the low-energy theorem of $\pi$, the $\pi\pi$ scattering amplitudes possess the Adler zero even for the $s$-wave, and therefore CDD pole contributions must exist.

Eq.~(\ref{eq:chiral unitary}) provides a general representation of the scattering amplitude based on its analytic structure. By unitarity, $G(s)$ is essentially fixed, whereas $V_\ell(s)$ encodes nontrivial information about the dynamics of the system and cannot be determined on general grounds, making the determination of $V_\ell(s)$ the central issue. We note that $T_\ell(s)$ can be written as
\begin{align}
    \label{eq:Kernel V}
    T_\ell(s) = V_\ell(s) - V_\ell(s)G(s)T_\ell(s),
\end{align}
and, as we will explain shortly, since $G(s)$ represents the loop effects from two-body scattering, this identity is equivalent to the Lippmann–Schwinger equation. Then, in the chiral unitary approach one identifies $V_\ell(s)$ with the Born amplitudes obtained from perturbation theory (e.g, ChPT), thereby incorporating the leading-order interaction into the resummed scattering amplitude. One may note that, despite its name, this procedure is not restricted to ChPT, as it relies only on the general analytic properties of the amplitude.

We next turn to the loop function $G(s)$, which represents the contribution from the two-body intermediate states and takes the following dispersion integral of $\rho(s)$:
\begin{align}
    \label{eq:loop function}
    G(s) = -\frac{1}{\pi}\int^\infty_{s_{\rm th}}ds' \frac{\rho(s')}{s'-s}.
\end{align}
Note that this representation is only formal, since the integral on the right-hand side diverges.
Accordingly one subtraction is needed in the dispersion relation, introducing a subtraction constant as a free parameter. Although, phenomenologically, this subtraction constant is arbitrary and treated as a model parameter, here we fix it using the so-called natural value estimation\,\cite{Oller:1998hw,Oller:2019opk,Oller:2000ma}. There is, in general, no guarantee that the phenomenological value of the subtraction constant obtained from model fitting coincides with its natural-value estimate. However, as we discuss in Sec.~\ref{sec:compositeness}, a discrepancy between the phenomenological and natural values of the subtraction constant has an important physical implication.

We introduce the natural value estimation, which begins with noting that Eq.~(\ref{eq:loop function}) is equivalent to the one-loop two-point diagram with external momentum $P$ satisfying $P^2=s$:
\begin{align}
    \label{eq:loop function diagram}
    G(s) 
    = \int \frac{d^4q}{(2\pi)^4}\frac{i}{[(P/2-q)^2-m_\pi^2+i0^+][(P/2+q)^2-m_\pi^2+i0^+]}.
\end{align}
One can explicitly verify the relation, by applying the Cutkosky rule $i(p^2-m_\pi^2+i0^+)^{-1} \to -2\pi i \delta(p^2-m_\pi^2)$ to the propagators in the integral to take the imaginary part of $G(s)$ and reconstructing it through the dispersion relation. While the imaginary part of the loop integral Eq.~(\ref{eq:loop function diagram}) remains finite, the real part exhibits a logarithmic divergence. This, of course, corresponds to the fact that the dispersion relation for the loop function $G(s)$ in Eq.~(\ref{eq:loop function}) requires one subtraction in order for the integral to converge.

We evaluate Eq.~(\ref{eq:loop function diagram}) both in dimensional regularization and in cutoff regularization:
in dimensional regularization, one obtains
\begin{align}
    \label{eq:loop function in DR}
    G(s) = \frac{1}{16 \pi^2}\left(a(\mu)+\ln \frac{m_\pi^2}{\mu^2} -\sigma(s) \ln \frac{\sigma(s)-1}{\sigma(s)+1}\right),
\end{align}
where the subtraction constant $a(\mu)$ is introduced and the divergent terms are absorbed into it. Here, $\mu$ is the ’t Hooft unit of mass, corresponding to the renormalization scale. It is introduced to render the arguments of logarithms dimensionless in dimensional regularization. From the viewpoint of dispersion relations, $\mu$ plays the role of the subtraction point, and therefore the final, physically observable result must be independent of $\mu$. We also define $\sigma(s)=16\pi \rho(s)$ and choose the branch cut of the logarithm to lie along the negative real axis.
In the cutoff regularization scheme, $G(s)$ is evaluated by introducing a momentum cutoff $\Lambda$ for the loop integral,
\begin{align}
    G_{\Lambda}(s) = \int^\Lambda_0 \frac{p^2 dp}{2\pi^2 E(p)}\frac{1}{s-4E(p)^2+i0^+},
\end{align}
where $E(p)=\sqrt{p^2+m_\pi^2}$.
The full analytic result of this integral is found in Ref.~\cite{Oller:1998hw} and its explicit form at the threshold is given by
\begin{align}
    G_{\Lambda}(s_{\rm th}) 
    =
    -\frac{1}{8\pi^2}
    \left[
        \ln
            \left(
                1+\sqrt{1+\frac{m_\pi^2}{\Lambda^2}}
            \right)
        -\ln \frac{m_\pi}{\Lambda}
    \right].
\end{align}
Since dimensional and cutoff regularizations differ only in the evaluation scheme, imposing the matching condition that they coincide at the threshold $s=s_{\rm th}$ yields the subtraction constant $a(\mu)$ in terms of the cutoff scale:
\begin{align}
    \label{eq:natural value}
    a(\mu) = -2\ln\left(1+\sqrt{1+\frac{m_\pi^2}{\Lambda^2}}\right)-\ln \frac{\Lambda^2}{\mu^2}.
\end{align}
Eq.~(\ref{eq:loop function in DR}) and Eq.~(\ref{eq:natural value}) show that the loop function $G(s)$ indeed does not depend on the renormalization scale $\mu$, as expected, while it now depends on the cutoff scale $\Lambda$. 

All the ingredients required for the chiral unitary method are now in place: the interaction kernel $V_\ell(s)$ is obtained from the perturbative calculation in ChPT, the subtraction constant is determined from the cutoff scale of ChPT through Eq.~(\ref{eq:natural value}), the loop function $G(s)$ is then evaluated from Eq.~(\ref{eq:loop function in DR}), and finally, an improved amplitude consistent with unitarity is obtained via Eq.~(\ref{eq:chiral unitary}). Furthermore, while the above discussion has been limited to elastic scattering, the same procedure can also be applied to coupled channels. The optical theorem for the $T$-matrix Eq.~(\ref{eq:optical theorem}) can be written in a form analogous to Eq.~(\ref{eq:optical inverse}), allowing one to construct a dispersion relation as in the elastic case. The only difference from the single-channel case is that both the kernel and the loop function are now matrices.

It should be stressed that the chiral unitary method is a model rather than an exact solution of QCD. In practice, the right-hand (unitarity) cut is treated nonperturbatively via resummation, whereas the left-hand cut and possible CDD poles are encoded in a finite-order polynomial constrained only by the low-energy chiral expansion. In particular, while the resulting unitarized amplitude implements exact two-body unitarity and the correct right-hand cut, it does not fully respect crossing symmetry, since the left-hand cut is only modeled approximately.

To conclude, we briefly comment on the so-called $K$-matrix method\,\cite{Newton:1982qc,Oller:2020guq}.
For $s > s_{\rm th}$, the logarithmic term in $G(s)$ develops an imaginary part, ${\rm Im}\, G(s + i0^+) = -\rho(s)$, so that the scattering amplitude $T_\ell(s)$ in Eq.~(\ref{eq:chiral unitary}) can be written as
\begin{align}
    T_\ell(s) = \big[K_\ell^{-1}(s) - i\rho(s) \big]^{-1},
\end{align}
where $K^{-1}_\ell(s) = V_\ell^{-1}(s) + {\rm Re}\,G(s)$ and $K_\ell(s)$ is referred to as the $K$-matrix. Since $K_\ell(s)$ is real above threshold, the $K$-matrix representation provides a parametrization of the scattering amplitude $T_\ell(s)$ using only real functions. The same procedure is also applicable in the coupled-channel case, with the $K$-matrix and the phase-space factor $\rho(s)$ promoted to matrices in channel space. One simplification sometimes used in the $K$-matrix approach is to neglect the real part of the loop function $G(s)$, approximating the $K$-matrix by the leading-order amplitude\,\cite{Aydemir:2012nz,Kamada:2022zwb}
\begin{align}
    \label{eq:K-matrix approximation}
    K_\ell(s) \simeq T_\ell^{\rm LO}(s).
\end{align}
This construction is a minimal deformation of the LO calculation to make it consistent with unitarity and is convenient, but it is unlikely to be effective for ChPT. Indeed, because it fixes the near-threshold amplitude to its LO value, it cannot reproduce the $I=0$ pion scattering length. To resolve this issue, the contribution of ${\rm Re}\,G(s)$ is essential.

\subsection{Application to QCD pion scattering}

Let us examine the improvement of the scattering amplitude achieved by the chiral unitary method in Eq.~(\ref{eq:chiral unitary}) with the natural value estimation of the subtraction constant Eq.~(\ref{eq:natural value}) for QCD pions, focusing on the $\ell=0$, $I=0$ $\pi\pi$ elastic channel.\footnote{The impact of the improvement using the natural value estimation on the $I=1$ and $I=2$ channels is small.}
In this case, the interaction kernel $V^0_0(s)$ is given by the LO chiral amplitude in Eq.~(\ref{eq:LO amplitudes}) as
\begin{align}
    V^0_0(s) \simeq \frac{s-m_\pi^2/2}{f_\pi^2}.
\end{align}
We fix the natural value of the subtraction constant to $a(\mu=1\,{\rm GeV}) \simeq -0.88$ by taking the cutoff as $\Lambda \simeq 770 \,\mathrm{MeV}$ in Eq.~(\ref{eq:natural value}). Here, the cutoff is set conservatively at the $\rho$-meson mass, beyond which the pion-only effective field theory ceases to be reliable.
With these inputs and Eq.~(\ref{eq:chiral unitary}), we obtain the improved amplitude as
\begin{align}
    \label{eq:chiral unitary amplitude real pi}
    T^0_0(s) = \left(\frac{f_\pi^2}{s - m_\pi^2/2} + G(s) \right)^{-1},
\end{align}
which incorporates the LO chiral amplitude as the interaction kernel and the rescattering effects through $G(s)$, providing a resummed amplitude that satisfies unitarity. Since the LO amplitude has an Adler zero, $V_0^0(s)^{-1}$ must include a corresponding CDD-pole term. In this matching, it is implemented as a simple pole at $s=m_\pi^2/2$ with residue $f_\pi^2$.

\begin{figure}[t]
\centering
\includegraphics[width=0.7\linewidth]{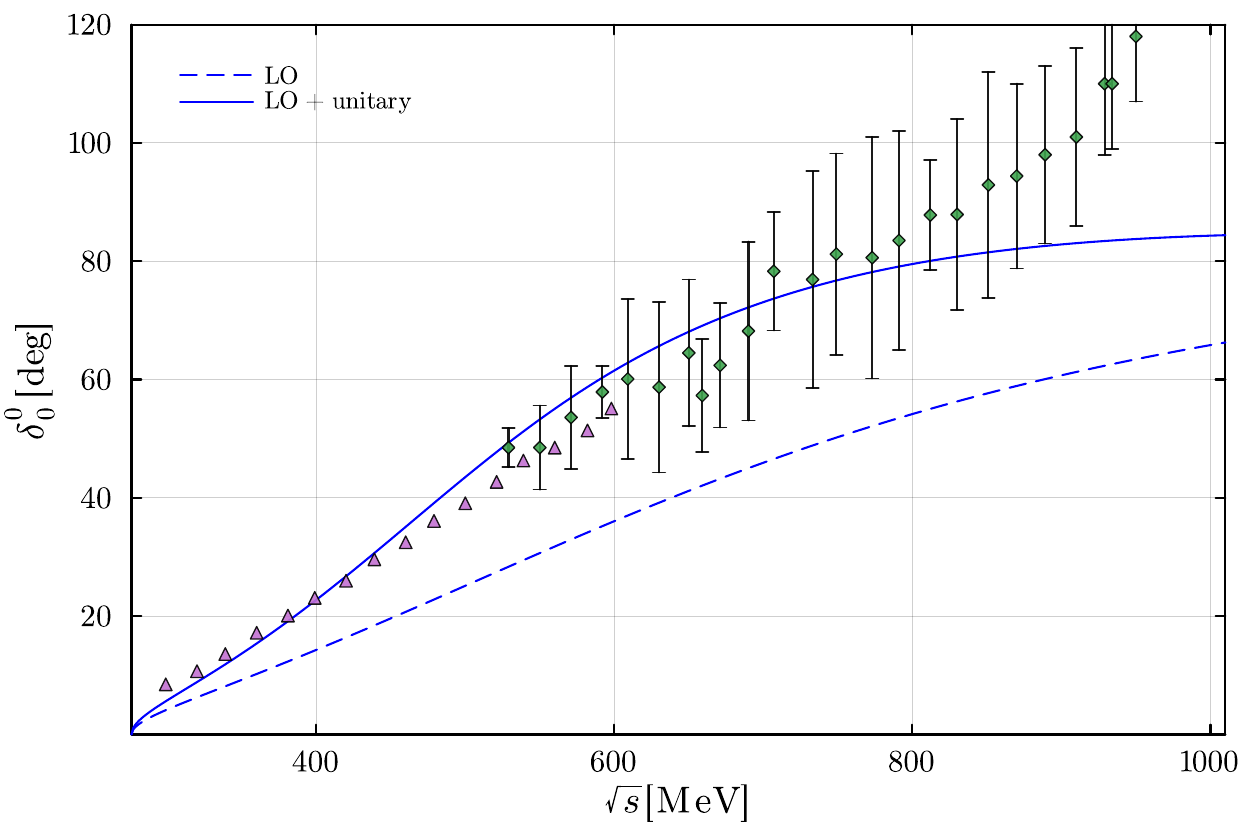}
    \caption{\small \sl
    Phase shift $\delta_0^0$ (in degrees) up to $\sqrt{s}=1\,{\rm GeV}$. The dashed curve is the LO result computed with the $K$-matrix approximation Eq.~(\ref{eq:K-matrix approximation}); the solid curve is the result from the chiral unitary method in Eq.~(\ref{eq:chiral unitary amplitude real pi}); and the triangle and diamond markers indicate values extracted from experimental data\,\cite{Mennessier:1982fk,Oller:2000ma}.
    }
    \label{fig: phaseshift}
\end{figure}

From Eq.~(\ref{eq:chiral unitary amplitude real pi}), the pion scattering length in the chiral unitary method is obtained as $a^0_0 \simeq 0.212$, which is much closer to the sophisticated value in Eq.~(\ref{eq:scattering length Roy}) than the prediction from the LO Weinberg result in Eq.~(\ref{eq:Weinberg}). Moreover, the improved amplitude reproduces the experimental data not only near threshold but also over a much broader kinematic range.
Fig.~\ref{fig: phaseshift} shows the phase shift $\delta^0_0$ (in degrees) up to $\sqrt{s}=1\,\mathrm{GeV}$: the dashed curve denotes the LO phase shift using the $K$-matrix approximation in Eq.~(\ref{eq:K-matrix approximation}); the solid curve shows the result of the chiral unitary method; and the triangle and diamond points represent the values extracted from experimental data\,\cite{Mennessier:1982fk,Oller:2000ma}. Since the tree-level amplitude does not satisfy unitarity, the phase shift is not globally well defined; therefore, as a simple remedy we present the LO result by the $K$-matrix form in Eq.~(\ref{eq:K-matrix approximation}).
While the LO calculation underestimates the phase shift throughout the entire energy region including threshold, the chiral unitary method reproduces the data well up to about $\sqrt{s}\sim 800 \,\mathrm{MeV}$.
Beyond this energy, a discrepancy gradually appears, as the present calculation is based on the $SU(2)\times SU(2)/SU(2)$ chiral perturbation theory where $K$ mesons are neglected, and thus the $f_0(980)$ resonance effect is not included. Incorporating these effects extends the agreement to higher energies\,\cite{Oller:1998hw,Oller:2000ma}. 

The primary reason for the failure of the LO calculation is the influence of the broad $\sigma$ resonance or $f_0(500)$.
The improvement achieved by the chiral unitary method properly incorporates this resonance.
By analytically continuing the loop function $G(s)$, defined on the physical sheet, to the second Riemann sheet or the unphysical sheet by adding its discontinuity across the cut, $G_{\mathrm{II}}(s)=G(s)+2i\rho(s)$, the amplitude itself is also analytically continued, allowing the corresponding resonance pole to be identified.
Numerically, we obtain $\sqrt{s_\sigma}=0.453-i\,0.238 \,\mathrm{GeV}$, which is consistent with the value reported by the Particle Data Group $\sqrt{s_\sigma} = (0.4-0.5) - i\,(0.20-0.35)\,{\rm GeV}$\,\cite{ParticleDataGroup:2024cfk} and sophisticated dispersive studies $\sqrt{s_\sigma} \simeq 0.45 -i\,0.250 \,{\rm GeV}$\,\cite{Garcia-Martin:2011nna}. Therefore, despite relying only on the LO perturbation and a cutoff, the chiral unitary amplitude with the natural-value estimation in Eq.~(\ref{eq:chiral unitary amplitude real pi}) offers a remarkably predictive framework consistent with experimental data.

\subsection{Variation of the amplitude with the coupling}
\label{subsec:mf-variation}

Having confirmed the importance of the chiral unitary method in the $I=0$ channel of QCD pions, we now turn to its impact on the dark pion system. In this case, since the coupling constant $m_\pi/f_\pi$ is an unknown parameter, we examine how the scattering amplitude behaves as this ratio is varied. To focus on the qualitative behavior of the amplitude, we assume the same symmetry-breaking pattern as in QCD, namely $SU(2)\times SU(2)/SU(2)$, and concentrate on the $I=0$ channel. For the subtraction constant, we adopt the natural-value estimation. Although the cutoff scale of dark-sector ChPT is also unknown, we fix it to the intrinsic chiral scale $\Lambda_\chi = 4\pi f_\pi$ for definiteness. A variation of the cutoff by an $\mathcal{O}(1)$ factor around $\Lambda_\chi$ changes $a(\mu)$ only by an $\mathcal{O}(1)$ amount, since the first term of Eq.~(\ref{eq:natural value}) is nearly constant $-2\ln 2$ up to $\mathcal{O}(m_\pi^2/\Lambda_\chi^2)$ when $m_\pi\!\ll\!\Lambda_\chi$, while $\Lambda\!\to\! c\,\Lambda$ shifts $a(\mu)$ by $\Delta a = -2\ln c$. From Eq.~(\ref{eq:loop function in DR}), this induces only an $s$–independent shift of the loop function $\Delta G(s) = \Delta a(\mu)/16\pi^2$ so the real part of $T_\ell^{-1}(s) = V_\ell^{-1}(s)+G(s)$ is modified merely by a numerically small constant. Consequently, the qualitative features of the unitarized amplitude—such as the presence or absence of a resonance—remain stable as long as the cutoff is taken around $\Lambda_\chi$, although quantitative details (e.g.\ pole positions or the scattering length) may shift at the level expected from an $\mathcal{O}(1)$ change in $a(\mu)$. Conversely, significantly changing the qualitative behavior of the amplitude would require pushing the cutoff far away from the chiral scale, i.e.\ taking $a(\mu)$ far from its natural value. We return to this point in Sec.~\ref{sec:compositeness}.

\begin{figure}[t]
    \centering
    \includegraphics[width=0.7\linewidth]{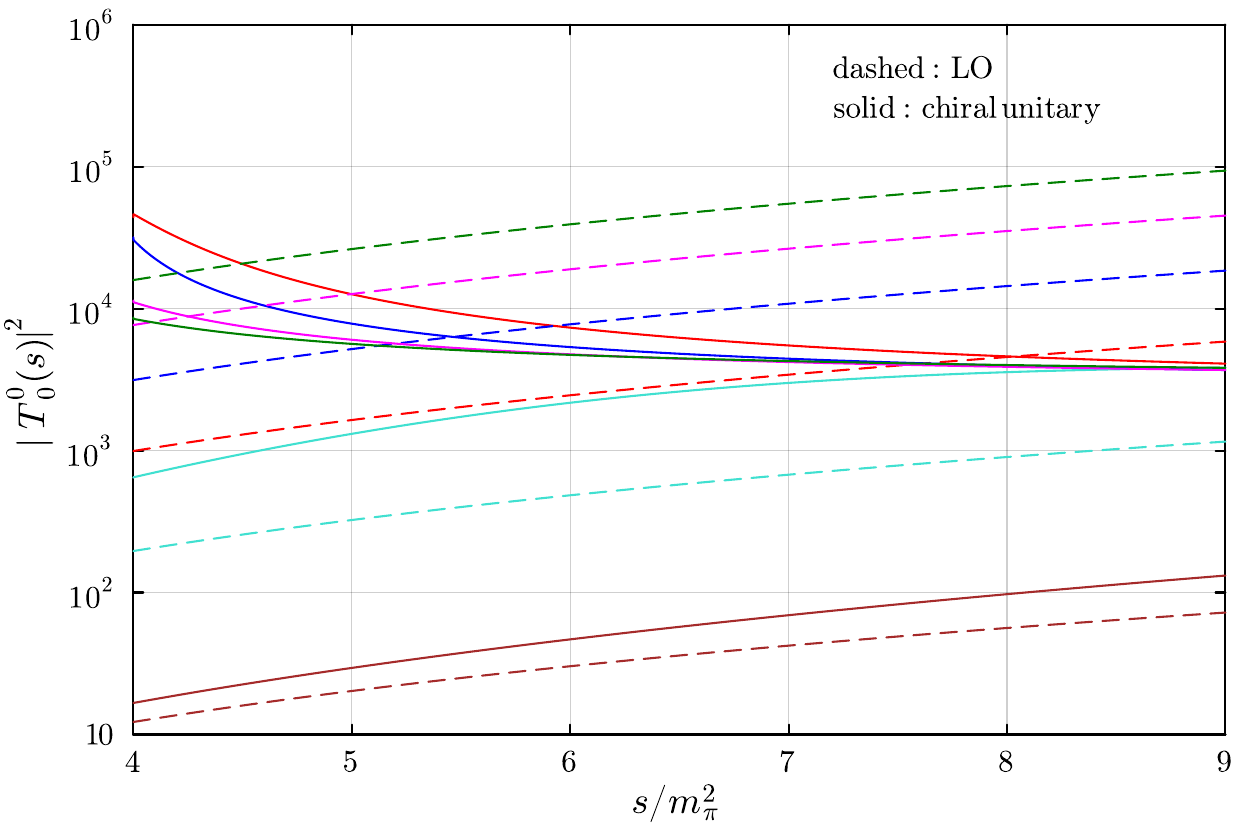}
    \caption{\small \sl 
    Squared amplitudes $|T_0^0(s)|^2$ obtained from the LO calculation (dashed) and from the chiral unitary method (solid) for couplings $m_\pi/f_\pi = 1,2,3,4,5,6$ (brown, cyan, red, blue, magenta, green).
    }
    \label{fig: su2multi}
\end{figure}

Fig.~\ref{fig: su2multi} shows the $s$-dependence of the amplitude squared $|T_0^0(s)|^2$ for various values of the coupling $m_\pi/f_\pi$. The dashed lines represent the LO amplitudes, while the solid lines correspond to the results obtained with the chiral unitary method. The color indicates different values of the coupling constant as $m_\pi/f_\pi = 1, 2, 3, 4, 5, 6$ for brown, cyan, red, blue, magenta, and green, respectively.
In the LO calculation, the amplitude squared increases with $m_\pi/f_\pi$ at all $s$, showing a monotonic enhancement. In contrast, the chiral unitary amplitudes exhibit qualitatively different behavior. For $m_\pi/f_\pi = 1,2$, they are $\mathcal{O}(1)$ larger than the LO results and remain monotonic functions of $s$, whereas for $m_\pi/f_\pi \ge 3$, the amplitudes develop a peak near the threshold $s = 4m_\pi^2$ and then decrease monotonically as $s$ increases, with the peak height gradually diminishing as the coupling becomes stronger.

\begin{figure}[t]
    \centering
    \includegraphics[width=0.7\linewidth]{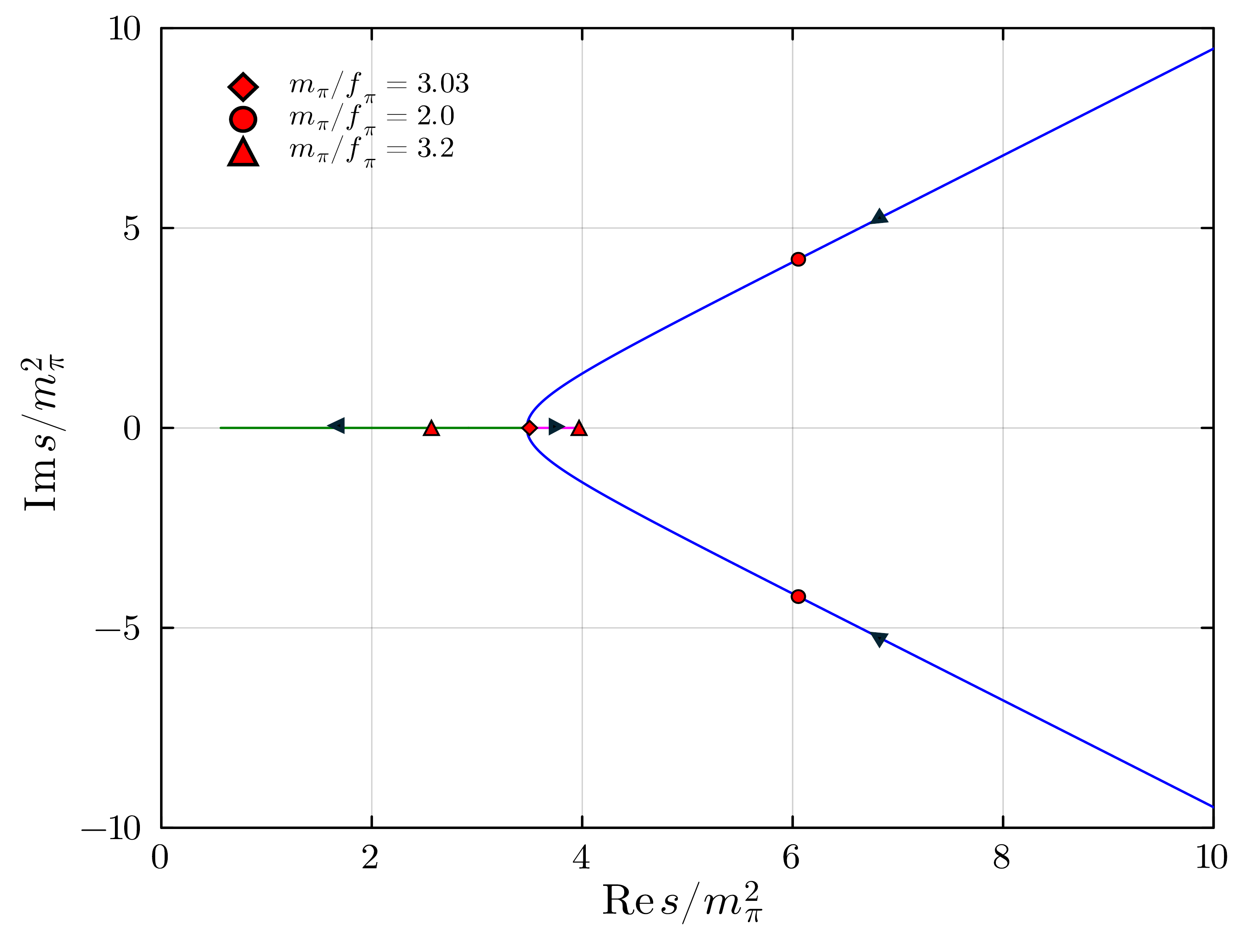}
    \caption{\small \sl
    Trajectory of the poles of the analytically continued $\ell=0$, $I=0$ scattering amplitude on the unphysical sheet as the coupling $m_\pi/f_\pi$ is varied from 1 to $4\pi$. 
    For couplings $m_\pi/f_\pi \lesssim 3$, a pair of resonance and anti-resonance poles exists (blue lines). As the coupling increases, the poles merge below the threshold on the real axis to form a virtual state. One of the poles then moves onto the negative real axis as a virtual-state pole (green line), while the other moves in the positive direction and eventually reaches the physical sheet, becoming a bound-state pole (magenta line).
    }
    \label{fig: poletrajectory}
\end{figure}

These behaviors are understood from the pole trajectories of the analytically continued amplitude on the unphysical sheet, shown in Fig.~\ref{fig: poletrajectory}, as the coupling constant is varied from $m_\pi/f_\pi = 1$ to $4\pi$.
For small couplings, $m_\pi/f_\pi \lesssim 3$, a pair of resonance and anti-resonance poles emerges on the complex plane, and both move toward the real axis as the coupling increases (blue lines).
At the critical value $m_\pi/f_\pi \simeq 3.03$, the two poles merge below the real-axis threshold and turn into a virtual state.
As the coupling increases further, the two poles move along the real axis in opposite directions (green and magenta lines); the pole moving toward positive $\mathrm{Re}\,s$ approaches the threshold from below at $m_\pi/f_\pi \simeq 3.2$. Beyond this point, it crosses onto the physical sheet and becomes a bound-state pole, while both poles continue to move leftward on the real axis as the coupling grows.

The pole trajectory with respect to the coupling $m_\pi/f_\pi$ explains the behavior of the amplitudes shown in Fig.~\ref{fig: su2multi}.
For small couplings, a broad resonance is present, leading to an enhancement of the amplitude relative to the LO result ($m_\pi/f_\pi = 1,2$).
As the coupling increases, a shallow bound or virtual state emerges, producing a resonant-like enhancement near threshold ($m_\pi/f_\pi = 3$).
For larger couplings, as the binding energy increases and the bound-state pole moves farther away from the threshold, the resonance peak diminishes and only its tail remains visible ($m_\pi/f_\pi = 4, 5, 6$).
These behaviors of the poles as a function of the coupling constant follow the general expectations from scattering theory\,\cite{Newton:1982qc} and are consistent with the results of more refined analyses, such as lattice-QCD studies of the coupling dependence of the $\sigma$ resonance\,\cite{Hanhart:2008mx,Pelaez:2015qba,Rodas:2023gma}.

\section{Implications for dark-pion DM
}
\label{sec:Implication}

We assess the extent to which amplitudes obtained via the chiral unitary method, relative to a LO calculation, can alter the phenomenological predictions of models in which dark pions constitute DM. As representative classes of dark-pion DM, we consider the strongly interacting massive particle (SIMP)\,\cite{Hochberg:2014dra} and weakly interacting massive particle (WIMP)\,\cite{Arcadi:2017kky} scenarios.
We consistently adopt the natural–value estimate in Eq.~(\ref{eq:natural value}) for the subtraction constant throughout this section, taking the cutoff at the chiral scale $\Lambda_\chi = 4\pi f_\pi$. This choice defines a minimal benchmark framework. Nevertheless, as described before, from a model-evaluation standpoint the subtraction constant is merely a phenomenological parameter. Indeed, a discrepancy between phenomenologically fitted subtraction constants and their natural estimates has been argued to correlate with the compositeness of resonances. We therefore discuss this aspect as well.

\subsection{SIMP case}

We begin by discussing the implications of applying the chiral unitary method to the case in which SIMP DM is realized by dark pions\,\cite{Hochberg:2014kqa,Hochberg:2015vrg,Hochberg:2018rjs,Choi:2016hid,Lee:2015gsa}. SIMP DM attains its relic abundance via freeze-out of number-changing $3\to2$ self-interactions. With $\mathcal{O}(1)$ strong coupling, the relic abundance condition naturally points its mass $m_{\rm DM}$ toward a sub-GeV (QCD-adjacent) scale—the “SIMP Miracle”\,\cite{Hochberg:2014dra}. Generally, $3\to2$ induces cannibalization, keeping the dark temperature above adiabatic cooling; but if elastic scattering with the SM maintains kinetic equilibrium, the evolution of the number density $n$ of SIMP DM reduces to the standard Boltzmann equation
\begin{align}
    \label{eq:SIMP Boltzmann}
    \frac{dn}{dt} + 3Hn  &= -\braket{\sigma_{3\to2} v^2}(n^3-n^2n_{\rm eq}),
\end{align}
where $H$ is the Hubble parameter and $n_{\rm eq}$ is the  equilibrium number density. $\sigma_{3\to2}v^2$ and $\braket{\sigma_{3\to2}v^2}$ denote the $3\to2$ cross section and thermal average of it, with $v$ the Jacobi velocity of the three-body system, defined by $s=9m_{\rm DM}^2+\sqrt{3}m_{\rm DM}^2v^2$\,\cite{Kamada:2022zwb}. When there are multiple DM species, the relevant cross sections must be averaged over the species—i.e., over all initial states—with the same convention applied to thermally averaged rates. The $3\to2$ cross section $\sigma_{3\to2}v^2$ is related, via the detailed balance relation, to the $2\to3$ cross section $\sigma_{2\to3}v_{\rm rel}$, where $v_{\rm rel}$ is the Møller velocity of the two-body system, as
\begin{align}
    \label{eq:detaile balance}
    \frac{(2m_{\rm DM})^3}{3!}\Phi_3(s)(\sigma_{3\to2}v^2) = \frac{(2E)^2}{2!}\Phi_2(s)(\sigma_{2\to3}v_{\rm rel}).
\end{align}
Here, $(2m_{\rm DM})^3$ and $(2E)^2$ encode the initial-state density factors, in which $E$ denotes the single-particle energy of DM.
The left-hand side uses the nonrelativistic approximation $E\simeq m_{\rm DM}$ for the three-body initial state, whereas the right-hand side retains the relativistic form at the same invariant energy $\sqrt{s}$.
$\Phi_2(s)$ and $\Phi_3(s)$ are the two- and three-body phase-space volumes, whose explicit forms are given by 
\begin{align}
\Phi_2(s) = \frac{1}{8\pi}\sqrt{1-\frac{4m_\pi^2}{s}},\quad\Phi_3(s) \simeq \frac{1}{192\sqrt{3}\pi^2}(\sqrt{s}-3m_{\rm DM})^2,
\end{align}
where for the three-body phase space we have employed the nonrelativistic approximation valid near threshold. The factorials $3!$ and $2!$ remove overcounting for identical initial particles.

The Boltzmann equation in Eq.~(\ref{eq:SIMP Boltzmann}) cannot be solved analytically, but a semi-analytic solution is known. By neglecting the second term on the right-hand side of Eq.~(\ref{eq:SIMP Boltzmann}) due to the Boltzmann suppression after the freeze-out temperature $T_f$, an approximate formula for the relic abundance of SIMP DM can be obtained as\,\cite{Choi:2017mkk}
\begin{align}
    \label{eq:SIMP approximation}
    \Omega_{\rm SIMP}h^2 &\simeq \frac{1.05 \times 10^{-10}\,{\rm GeV}^{-5}}{g_*^{3/4}(T_f)(K(x_f)/M_{\rm pl})^{1/2}}, \quad 
    K(x_f) = \int^\infty_{x_f} \frac{dx}{x^5} \braket{\sigma_{3 \to 2} v^2},
\end{align}
where, $g_*(T_f)$ is the effective degrees of freedom at freeze-out, $x_f = m_{\rm DM} / T_f$, and $M_{\rm pl}=2.4\times10^{18}\,{\rm GeV}$ is the reduced Planck mass. Numerically, $x_f$ is typically in the range of $x_f=14$ to $24$\,\cite{Hochberg:2014dra}. 

One of the attractive features of the SIMP scenario is its self-interaction: SIMP DM that satisfies the relic abundance condition $\Omega_{\rm SIMP} h^2 \simeq 0.12$ naturally—or even inevitably—leads to a sizable elastic self-scattering cross section $\sigma_{\rm DM}$. While too large a self-interaction of DM is constrained by observations, it has been suggested that a self-interaction of moderate strength could resolve small-scale structure problems\,\cite{Spergel:1999mh,Tulin:2017ara}. Strictly speaking, the constraints depend on which galactic or cluster system is considered and on the corresponding characteristic DM velocity. Since our goal here is to assess the impact of unitarization on self-scattering, for simplicity, we adopt the commonly used observationally favorable range
\begin{align}
    \label{eq:self-scattering condition}
    0.1\,{\rm cm^2/g} \lesssim \frac{\sigma_{\rm DM}}{m_{\rm DM}} \lesssim 1.0\,{\rm cm^2/g},
\end{align}
where the upper bound comes from the Bullet cluster\,\cite{Kahlhoefer_2013} while the lower bound is required to address the core–cusp problem\,\cite{Zavala_2013}. In the following discussion, we evaluate the self-scattering cross section at a relative DM velocity $v_{\rm DM}=10^{-3}$, which is representative of galactic-scale halos.

Before we consider the impact of the chiral unitary method on dark-pion SIMPs, let us briefly outline the partial-wave analysis in the presence of an inelastic $2\to3$ process\,\cite{Kamada:2022zwb}. Since a full partial-wave expansion of the $2\to3$ invariant amplitude, including the three-body final state, is technically involved, we project it onto the initial partial wave $\ell$ and absorb the three-body phase space into its normalization. Mimicking the elastic case in Eq.~(\ref{eq:elastic cross section partial wave}), the total $2\to3$ cross section is then written as
\begin{align}
    \sigma_{2\to3}(s) = \sum_{\ell=0}^\infty\frac{2\ell+1}{8\pi s}|T_{\ell,2\to3}(s)|^2,
\end{align}
where $T_{\ell,2\to3}(s)$ denotes the inelastic $2\to3$ partial-wave amplitude with the three-body phase-space factor absorbed. With this normalization of the inelastic partial-wave amplitude, the optical theorem Eq.~(\ref{eq:optical theorem}) extends to include the inelastic $2\to3$ channel as
\begin{align}
    \label{eq:optical theorem inelastic}
    {\rm Im}\,T_{\ell,2\to2} (s) = \rho(s) (|T_{\ell,2\to2} (s)|^2+|T_{\ell,2\to3}(s)|^2).
\end{align}
The unitarity of the $S$-matrix extends to inelastic channels, so there are also the optical-theorem relations for the $2\to3$, $3\to2$ and $3\to3$ amplitudes, $T_{\ell ,2\to 3}(s)$, $T_{\ell,3\to 2}(s)$ and $T_{\ell,3\to3}(s)$. They are given by
\begin{align}
    \label{eq:Watson}
    {\rm Im}\,T_{\ell,2\to3}(s) \simeq \rho(s) T_{\ell,2\to2}(s)T_{\ell,2\to3}(s)^*, \quad
    {\rm Im}\,T_{\ell,3\to3}(s) \simeq \rho(s)|T_{\ell,2\to3}(s)|^2.
\end{align}
Here we assume time-reversal invariance, so $T_{\ell,3\to2}(s) = T_{\ell,2\to3}(s)$, and we neglect the contribution of $T_{\ell,3\to3}(s)$ on the right-hand sides. Eqs.~(\ref{eq:optical theorem inelastic}) and (\ref{eq:Watson}) can be collectively written in a form analogous to Eq.~(\ref{eq:optical inverse}) as
\begin{align}
    {\rm Im}\,T_\ell(s)^{-1} \simeq -\rho(s), 
    \quad
    T_\ell(s)
    =
    \begin{pmatrix}
        T_{\ell,2\to2}(s) & T_{\ell,2\to 3}(s) \\
        T_{\ell,2\to 3}(s) & T_{\ell,3\to3}(s)
    \end{pmatrix},
\end{align}
where $T_\ell(s)$ is the $T$-matrix of $\pi\pi-\pi\pi\pi$ system. The above discussion applies, when the SIMP carries an internal symmetry, separately to initial states in each irreducible representation.

The optical theorem in Eq.~(\ref{eq:optical theorem inelastic}) places an upper bound on the $2\to3$ inelastic partial-wave amplitude, $|T_{\ell,2\to3}(s)| \le 1/4\rho(s)^2 $. Through the detailed-balance relation Eq.~(\ref{eq:detaile balance}) and assuming that only a finite number of partial waves contribute to the $3\to2$ channel, this bound propagates to $3\to2$ cross section and to its thermal average, implying the unitarity bound of the thermal average $\braket{\sigma_{3\to2}v^2}_{\rm uni} \propto m_{\rm DM}^{-5}$ in the nonrelativistic regime \cite{Kamada:2022zwb,Kuflik:2017iqs,Namjoo:2018oyn,Bhatia:2020itt}. It then follows from Eq.~(\ref{eq:SIMP approximation}) that the relic abundance attainable in the SIMP scenario is bounded from below by a function of the DM mass; hence, reproducing the observed value $\Omega_{\rm DM}h^2 \simeq 0.12$ imposes an upper bound on the viable SIMP DM mass of order $\mathcal{O}(1\,{\rm GeV})$.

We now consider a dark-pion realization of SIMP DM. The number-changing process $\pi\pi\pi\to\pi\pi$ is supplied by the Wess–Zumino–Witten (WZW) term\,\cite{Wess:1971yu,Witten:1979vv}: if the underlying QCD-like gauge theory carries a global anomaly, anomaly matching requires that the same anomaly appear in the low-energy effective theory of the dark pions, i.e., ChPT. Since the anomaly is topological, in ChPT it appears as the topological WZW term, written at the Lagrangian level as follows:
\begin{align}
    \mathcal{L}_{\rm WZW} = \frac{2k}{15\pi^2f_\pi^5}\epsilon^{\mu \nu \rho \sigma}{\rm Tr}\,\big[\pi\partial_\mu \pi \partial_\nu \pi \partial_\rho \pi \partial_\sigma \pi \big],
\end{align}
where level $k$ is an integer determined by matching to the UV anomaly and $\pi = \pi^aT^a$ with $T^a$ the broken generators of $G/H$. The $2\to3$ process generated by the WZW term is necessarily $p$-wave, as the five-point vertex carries four derivatives, making the amplitude of the irreducible representation $R$, $T^R_{\ell,2\to3}(s)$, linear in the initial relative momentum. By time-reversal symmetry, the inverse $3\to2$ process is likewise $p$-wave. Furthermore, only antisymmetric irreducible representations $R$ contribute: the two identical bosons require an overall symmetric state, and a $p$-wave has an antisymmetric spatial part that must be compensated by an antisymmetric internal one.

We now examine how improving the amplitudes with the chiral unitary method modifies the phenomenology of dark-pion SIMP DM. Unitarization can matter in two sectors: (i) the $p$-wave $3\to2$ process that sets the freeze-out, and (ii) the $s$-wave $2\to2$ self-scattering. The latter is straightforward, since both the initial and final states are two-body: for each irreducible representation $R$ that carries an $s$-wave component, we apply the chiral unitary method Eq.~(\ref{eq:chiral unitary}) to the LO partial-wave amplitudes. By contrast, a direct application of the same (single- or coupled-channel) machinery to $3\to2$ is obstructed by the presence of three-body states. In particular, two-body subsystems can develop $s$-wave resonances, and vector-meson–like excitations not explicit at LO ChPT may participate in unitarizing the amplitude\,\cite{Choi:2018iit}. We therefore do not attempt a full unitarization of the $3\to2$ amplitude and instead, following Ref.~\cite{Kamada:2022zwb}, we focus only on the previously stated unitarity bound for $3\to2$, and for the $p$-wave channels we merely employ a simple tree-level approximation as Eq.~(\ref{eq:K-matrix approximation}) for the $K$-matrix $K^R_1(s)$
\begin{align}
    \label{eq:inelastic K-matrix}
    K_{1}^R(s) 
    \simeq 
    \begin{pmatrix}
        T_{1,2\to2}^{R,\rm LO}(s) & T_{1,2\to 3}^{R,\rm LO}(s) \\
        T_{1,2\to 3}^{R,\rm LO}(s) & 0
    \end{pmatrix},
\end{align}
which leads to the following improved amplitudes
\begin{align}
    T_{1,2\to2}^{R}
    \simeq 
    \frac{T_{1,2\to2}^{R,\rm LO}+i\rho T_{1,2\to3}^{R,\rm LO}}{1-i\rho\big(T_{1,2\to2}^{R,\rm LO}+i\rho{T_{1,2\to3}^{R,\rm LO }}^2\big)},
    \quad
    T^R_{1,2\to3}
    \simeq 
    \frac{T^{R,\rm LO}_{1,2\to3}}{1-i\rho\big(T_{1,2\to2}^{R,\rm LO}+i\rho{T_{1,2\to3}^{R,\rm LO}}^2\big)}.
\end{align}
These two treatments—namely, the chiral unitary method applied to the $s$-wave channels and the tree-level $K$-matrix approximation adopted for the $p$-wave channels—are compatible, as freeze-out and self-scattering are governed by different dominant scattering channels. Accordingly, we employ channel-dependent unitarization schemes that provide a consistent and practical approximation for each observable within the same effective theory. Then our primary aim is to examine how unitarization impacts the self-scattering cross section implied by the coupling $m_\pi/f_\pi$ that satisfies the relic-abundance condition.

\begin{figure}[t] 
    \centering
    \includegraphics[width=\linewidth]{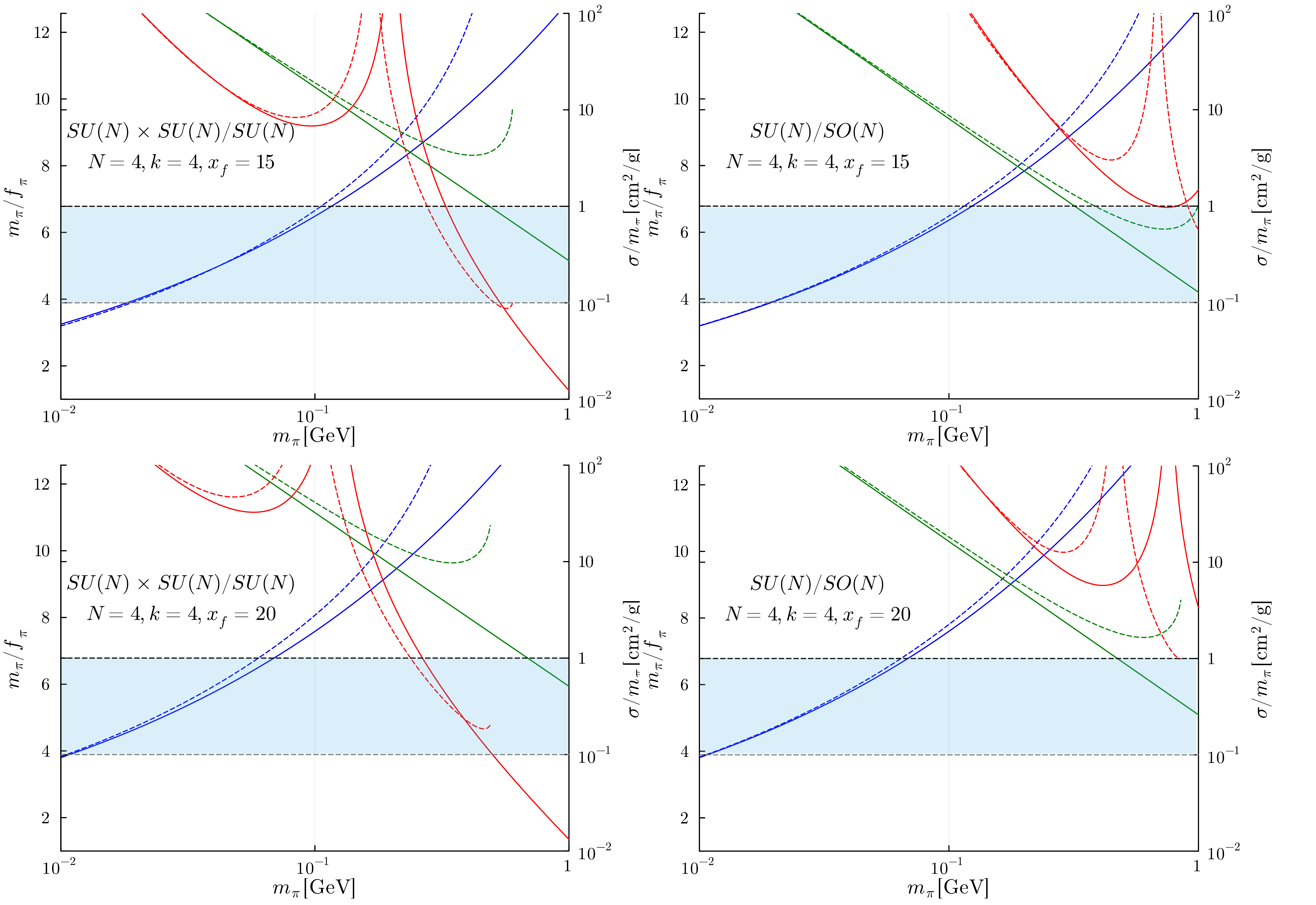}
    \caption{\small \sl  
    Couplings $m_\pi/f_\pi$ (left axis) that reproduce $\Omega_{\rm DM}h^2\simeq0.12$ under Eq.~(\ref{eq:SIMP approximation}) versus the dark-pion mass $m_\pi$, together with the corresponding self-scattering cross section $\sigma/m_\pi$ (right axis), evaluated at a representative galactic velocity of DM, $v_{\rm DM} = 10^{-3}$. Left panels: $SU(4)\times SU(4)/SU(4)$; right panels: $SU(4)/SO(4)$. Top: $x_f=15$; bottom: $x_f=20$. The phenomenologically preferred interval for $\sigma/m_\pi$ in Eq.~(\ref{eq:self-scattering condition}), $0.1\,{\rm cm^2/g} \,\lesssim\, \sigma_{\rm DM}/m_{\rm DM} \,\lesssim\, 1.0\,{\rm cm^2/g}$, is shown as a shaded band. Blue solid (dashed): couplings from LO (tree-level $K$-matrix) relic-abundance matching. Green solid/dashed: $\sigma/m_\pi$ evaluated at LO $2\to2$ using those couplings without/with unitarization of the freeze-out $3\to2$ process. Red solid/dashed: $\sigma/m_\pi$ computed with the $2\to2$ amplitude improved by the chiral unitary method in Eq.~(\ref{eq:chiral unitary}), based on couplings obtained without/with $3\to2$ unitarization.
    }
    \label{fig:wide}
\end{figure}

In Fig.~\ref{fig:wide}, we display the couplings $m_\pi/f_\pi$ (left axis) that reproduce the relic abundance $\Omega_{\rm DM}h^2 \simeq 0.12$ under the approximate formula Eq.~(\ref{eq:SIMP approximation}) for the given dark-pion DM mass $m_\pi$, along with the resulting self-scattering cross section $\sigma/m_\pi$ (right axis). Here we evaluate the unitarized $\sigma/m_\pi$ at $v_{\rm DM}=10^{-3}$, which we take as a representative velocity of DM on galactic scales. The left panels take the symmetry-breaking pattern of the dark pion to be $SU(4) \times SU(4)/SU(4)$, whereas the right panels take $SU(4)/SO(4)$. The upper panels use a freeze-out parameter $x_f=15$, and the lower panels use $x_f=20$. We also indicate in the figure the phenomenologically preferred interval for the self-scattering cross section, as given by Eq.~(\ref{eq:self-scattering condition}). In each panel, the blue solid (dashed) curves show, as a function of $m_\pi$, the couplings $m_\pi/f_\pi$ required to reproduce the relic abundance, computed at LO (with a tree-level $K$-matrix approximation). The green solid/dashed curves give the corresponding self-scattering cross section $\sigma/m_\pi$ evaluated at LO $2\to2$ amplitudes from those parameter choices $(m_\pi,m_\pi/f_\pi)$ without/with unitarization of the freeze-out $3\to2$ process. Finally, the red solid/dashed curves show $\sigma/m_\pi$ obtained using the improved $2\to2$ amplitude by the chiral unitary method in Eq.~(\ref{eq:chiral unitary}), based on the couplings without/with unitarization of the $3\to2$ process, respectively.

Using pure LO perturbation theory, all panels exhibit a parameter region of $(m_\pi,m_\pi/f_\pi)$ in which the required coupling stays below the naive perturbative bound, $m_\pi/f_\pi \lesssim 4\pi$, while the self-scattering cross section $\sigma/m_\pi$ falls within the phenomenologically favored window in Eq.~(\ref{eq:self-scattering condition}). On the other hand, when the $p$-wave amplitudes are unitarized via the tree-level $K$-matrix in Eq.~(\ref{eq:inelastic K-matrix}), the $3\to2$ inelastic amplitude is suppressed near the three-body threshold relative to the LO amplitude. To satisfy the relic-abundance condition at the same mass $m_\pi$, this requires a larger coupling than at LO calculation, which (i) lowers the mass upper limit imposed by the naive perturbative bound, and (ii) tends to inflate $\sigma/m_\pi$, making it more likely to fall outside the self-scattering window. 

In contrast, the self-scattering cross section $\sigma/m_\pi$ is strongly modified by the chiral unitary method in Eq.~(\ref{eq:chiral unitary}), and a resonance structure that depends on the symmetry-breaking pattern $G/H$—absent at LO—emerges. In both cases, a first prominent peak appears for couplings around $m_\pi/f_\pi \sim 4$. In addition, for the $SU(4) \times SU(4)/SU(4)$ cases a second peak occurs near $m_\pi/f_\pi \sim 8$, whereas for the $SU(4)/SO(4)$ cases it appears around $m_\pi/f_\pi \sim 12$. The seeds of these behaviors were already discussed in Sec.~(\ref{subsec:mf-variation}): the presence of near-threshold resonance poles. Although the self-scattering cross section $\sigma/m_\pi$ receives contributions from multiple irreducible representations, each can develop a dynamical resonance near threshold at appropriate values of the coupling $m_\pi/f_\pi$. As a result, the nonrelativistic $\pi \pi$ self-scattering is significantly shaped by these resonances. In the $SU(4) \times SU(4)/SU(4)$ case, when the relic density is matched with the LO $3\to2$ rate, the mass range satisfying the self-scattering window shrinks relative to using the LO self-scattering; if instead one matches with the unitarized $3\to2$ rate, a previously absent mass interval now falls inside the window. For the $SU(4)/SO(4)$ case, the mass range that satisfied the self-scattering window before adopting the chiral unitary method vanishes once the method is applied.

\begin{figure}[t] 
    \centering
    \includegraphics[width=0.7\linewidth]{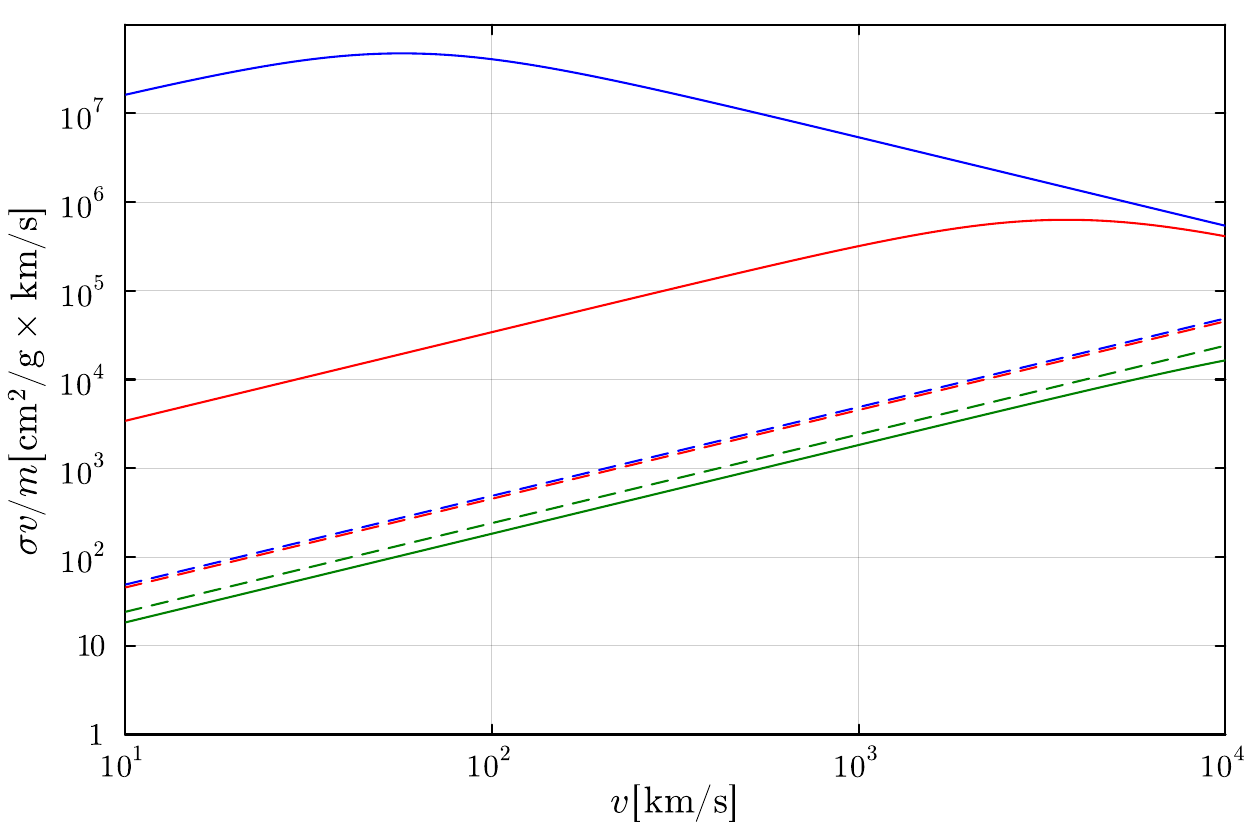}
    \caption{\small \sl  
    Self-scattering cross section times velocity of dark pion DM as a function of the dark pion velocity (in units of ${\rm km/s}$) for the symmetry-breaking pattern $SU(4)\times SU(4)/SU(4)$. The vertical axis is given in units of ${\rm cm^2/g}\times{\rm km/s}$, while we use the natural units everywhere else. The benchmark points $(m_\pi, m_\pi/f_\pi)$ are chosen as $(0.20\,{\rm GeV},8.0)$ (blue), $(0.21\,{\rm GeV},8.1)$ (red), and $(0.30\,{\rm GeV},9.0)$ (green). Dashed lines show the tree-level results, while solid lines correspond to the unitarized results. For the symmetry-breaking pattern $SU(4)\times SU(4)/SU(4)$, the zero-energy resonance appears at $m_\pi/f_\pi \sim 8$, so the benchmark points represent parameters near the resonance, slightly away from it, and well off resonance.
    }
    \label{fig:velocity dependence}
\end{figure}

We must note that the assessment of the impact of unitarization on the self-scattering cross section, summarized in Fig.\,\ref{fig:wide}, is evaluated for dark pion DM at a velocity $v_{\rm DM}=10^{-3}$. This is because, while the tree-level analysis indeed shows that the $s$-wave cross section does not exhibit any nontrivial velocity dependence at low energies, the unitarized cross section need not share this property, particularly when dynamically generated resonances appear near the threshold. Therefore, as an illustrative example, we consider the case where the symmetry breaking pattern is $SU(4)\times SU(4)/SU(4)$ (corresponding to the situation shown in the left panels of Fig.\,\ref{fig:wide}), and examine the velocity dependence of the cross section to conclude this section.

In Fig.\,\ref{fig:velocity dependence}, we show the self-scattering cross section times velocity of dark pion DM, plotted as a function of the dark pion velocity (in units of ${\rm km/s}$). The vertical axis is given in units of ${\rm cm^2/g}\times{\rm km/s}$. As benchmark points, the mass and coupling constant of the dark pion DM, ($m_\pi$,$m_\pi/f_\pi$), are chosen as $(0.20\,{\rm GeV},8.0)$ (blue line), $(0.21\,{\rm GeV},8.1)$ (red line), and $(0.30\,{\rm GeV},9.0)$ (green line). The tree-level results are shown by dashed lines, while the unitarized results are shown by solid lines. Since the zero-energy resonance for the symmetry-breaking pattern $SU(4)\times SU(4)/SU(4)$ appears at a coupling $m_\pi/f_\pi \sim 8$, these benchmark points are chosen to represent parameters near the resonance, slightly away from the resonance, and well off resonance. From the figure, we see that for the parameter set near the resonance (blue line), the scattering cross section exhibits a resonant structure at low velocities, i.e., a clear velocity dependence. For the red line, which corresponds to parameters somewhat away from the resonance, a deviation from the tree-level result is still present, although the peak position shifts to higher velocities.\,\footnote{In the approximation where the effective range is neglected at low energies, the self-scattering cross section exhibits a peak at $v_{\rm DM} = 1/m_\pi|A|$, where $A$ denotes the scattering length\,\cite{Chu:2019awd}. The presence of a zero-energy resonance corresponds to a divergence of the scattering length, $A=\infty$. Therefore, coupling constants closer to the zero-energy resonance lead to a peak in the cross section at smaller velocities.
} In contrast, for the green line, which lies sufficiently far from the resonance, the unitarized self-scattering cross section no longer shows a noticeable velocity dependence in the low-energy region.

\subsection{WIMP case}

Next we discuss the case where dark pions realize a WIMP\,\cite{Bhattacharya:2013kma,Harigaya:2016rwr,Cheng:2021kjg,Abe:2024mwa,Arina:2019tib} and the implication of the chiral unitary method. Here, WIMP denotes DM whose relic abundance is set by freeze-out of number-depleting annihilations into SM states, ${\rm DM} + {\rm DM} \to {\rm SMs}$; with weak-scale coupling constants, the observed relic requirement naturally points to an electroweak-adjacent mass scale—the “WIMP Miracle”\,\cite{Lee:1977ua,Steigman:2012nb}.
Kinetic equilibrium with the SM is typically maintained through sizable interaction among them, and the number density obeys the standard Boltzmann equation
\begin{align}
    \label{eq:Boltzmann WIMP}
    \frac{dn}{dt} + 3Hn  &= -\braket{\sigma^{\rm ann} v_{\rm rel}}(n^2-n_{\rm eq}^2) ,
\end{align}
where $\braket{\sigma^{\rm ann}v_{\rm rel}}$ is the thermal average of the pair annihilation cross section for WIMP. The behavior of the relic abundance derived from the Boltzmann equation, Eq.~(\ref{eq:Boltzmann WIMP}), depends on whether DM is relativistic or non-relativistic at the time it decouples from the SM thermal bath. In the following, we focus on the case where DM decouples in the non-relativistic regime, for which the DM phase-space distribution is well described by the Maxwell--Boltzmann distribution. The thermal average of the annihilation cross section is therefore obtained by weighting with the Maxwell--Boltzmann distribution:
\begin{align}
\braket{\sigma^{\rm ann}v_{\rm rel}} = \int^\infty_0 dv_{\rm rel}\, (\sigma^{\rm ann}v_{\rm rel}) \,f_{\rm MB}(v_{\rm rel}),\quad f_{\rm MB}(v_{\rm rel}) = \frac{x^{3/2}}{2\sqrt{\pi}}v_{\rm rel}^2 \exp \left(-\frac{xv_{\rm rel}^2}{4}\right),
\end{align}
where $f_{\rm MB}(v_{\rm rel})$ denotes the Maxwell--Boltzmann distribution of the relative velocity.
The parameter $x$ characterizes $f_{\rm MB}(v_{\rm rel})$ and, in the context of thermal averaging, is identified as $x = m_{\rm DM}/T$.
The mean squared velocity derived from the Maxwell--Boltzmann distribution is then given by $\braket {v_{\rm rel}^2} = 6/x$, corresponding to a typical velocity $\sqrt{\langle v_{\rm rel}^2 \rangle} = \sqrt{6/x}$.

Although an analytic solution to the Boltzmann equation for WIMP DM in Eq.~(\ref{eq:Boltzmann WIMP}) is not known, as in the SIMP case, it can be solved semi-analytically by introducing a freeze-out parameter $x_f$. The resulting relic abundance is given by\,\cite{Choi:2017mkk}
\begin{align}
    \label{eq:relic WIMP}
    \Omega_{\rm WIMP}h^2 & \simeq \frac{8.53 \times 10^{-11}\,{\rm GeV}^{-2}}{g_*^{1/2}(x_f)J(x_f)},\quad J(x_f) = \int^\infty_{x_f} \frac{dx}{x^2} \braket{\sigma^{\rm ann} v_{\rm rel}}.
\end{align}
Numerically, the freeze-out parameter typically lies in the range $x_f \simeq 20$ to $30$. The formula in Eq.~(\ref{eq:relic WIMP}) tells us that at a qualitative level the relic abundance of DM scales approximately inversely with its pair-annihilation cross section, $\Omega_{\rm WIMP}h^2 \sim \braket{\sigma^{\rm ann}v_{\rm rel}}^{-1}$, which is intuitive: a larger cross section leaves fewer particles at late times. More quantitatively, the final abundance is set by the post-freeze-out integral $J(x_f)$, which encodes the cumulative depletion of the comoving number density from residual annihilations after decoupling; hence it is the magnitude of $J(x_f)$ that primarily determines $\Omega_{\rm WIMP}h^2$. In particular, when the annihilation cross section $(\sigma^{\rm ann}v_{\rm rel})$ is dominated by an $s$-wave contribution with only mild velocity dependence, Eq.~(\ref{eq:relic WIMP}) leads to the well-known canonical value of the annihilation cross section for WIMP DM,
\begin{align}
\label{eq:WIMP canonical}
(\sigma^{\rm ann}v_{\rm rel})_{\rm WIMP} \simeq 2.0 \times 10^{-26}\,{\rm cm^3/s},
\end{align}
where we have fixed the freeze-out parameter to $x_f=23$ in deriving this estimate.

We examine how improving the dark-pion WIMP self-scattering amplitude via the chiral unitary method can affect WIMP phenomenology. For simplicity, we assume that pair annihilation proceeds exclusively through the singlet representation of the dark-pion system. The key ingredient is the optical theorem in a coupled-channel setting in Eq.~(\ref{eq:optical theorem}): letting $X$ denote a SM species that interacts with the dark pion $\pi$, the optical theorem for the $\pi-X$ system takes the form:
\begin{align}
    \label{eq:optical WIMP}
    {\rm Im}\,T_{\ell,\pi \pi}(s) \simeq \rho_\pi(s)|T_{\pi \pi}(s)|^2, \quad {\rm Im}\,T_{\ell,\pi X}(s) \simeq \rho_{\pi}(s) T_{\ell,\pi \pi}(s) T_{\ell,\pi X}(s)^*,
\end{align}
where $\rho_\pi(s)$, $T_{\ell,\pi \pi}(s)$ and $T_{\ell,\pi X}(s)$ denote the phase-space factor of $\pi$, the $\ell$-th self-scattering amplitude $\pi \pi \to \pi \pi$ and the annihilation amplitude $\pi \pi \to XX$, respectively. Here, under the assumption that the $\pi X$ interaction is weak, we neglect the contribution to the imaginary part of the amplitude arising from the $XX$ intermediate state. In this approximation, the first relation reduces to the standard optical theorem for elastic scattering in Eq.~(\ref{eq:optical theorem}), whereas the second becomes crucial and is known as Watson’s theorem\,\cite{Oller:2019opk,Oller:2020guq,Watson:1952ji,ParticleDataGroup:2024cfk,Kamada:2023iol}: since the left-hand side is real, the right-hand side must also be real, implying that the annihilation amplitude $T_{\ell,\pi X}(s)$ carries the same phase shift as the self-scattering amplitude $T_{\ell, \pi \pi}(s)$. Consequently, the annihilation amplitude must inherit the analytic structure of the elastic channel. Physically, this means that initial-state rescattering due to self-interactions is inevitably present prior to annihilation —an effect known as the initial-state interaction (ISI)\,\cite{Oller:2019opk, Oller:2020guq}.

When the DM self-interaction is represented by a long-range interaction—such as that supplied by light mediator exchange—the ISI is widely known as the Sommerfeld effect\,\cite{Hisano:2002fk,Hisano:2003ec,Hisano:2004ds,Arkani-Hamed:2008hhe}. In this case, the analytic structure of the annihilation amplitude asserted by Watson’s theorem—namely, the one inherited from elastic scattering—can be incorporated by recasting the long-range force as a potential-scattering problem in nonrelativistic quantum mechanics and using the corresponding Jost function $\mathscr{J}_\ell(s)$\,\cite{Newton:1982qc,Hyodo:2020czb} to account for it, as $T_{\ell,\pi X}(s) = T^{\rm LO}_{\ell,\pi X}(s)/\mathscr{J}_\ell(s)$\,\cite{Watanabe:2025kgw}. The correction by the Jost function can become significant when the DM pair is sufficiently nonrelativistic. As a canonical example, the Jost function for Coulomb scattering is known analytically; accordingly, the electromagnetic interaction enhances the $\ell$-th electron–positron pair-annihilation cross section to the photons, for relative momentum $p=|\vec{p}|$, by the Sommerfeld–Sakharov factor
\begin{align}
    \label{eq:SE Coulomb}
    S_{\ell}(p)
    =
    \frac{2\pi \eta}{1-e^{-2\pi \eta}}\prod_{b=1}^{\ell}
    \left(
    1+\frac{\eta^2}{b^2}
    \right),
\end{align}
with the Sommerfeld parameter $\eta = m_e\alpha/2p$, where $m_e$ is the electron mass and $\alpha$ the fine structure constant\,\cite{Cassel:2009wt,Iengo:2009ni}. In general, the zeros of the Jost function correspond to composite states—bound states and resonances—generated by the long-range interaction. Accordingly, the DM annihilation amplitude becomes resonant, and the cross section can be strongly enhanced when $\sqrt{s}$ approaches the corresponding energies\,\cite{Hisano:2002fk,Hisano:2003ec,Hisano:2004ds,Beneke:2024iev,Watanabe:2025kgw}.

Given that ISI from weak long-range interactions is important in the nonrelativistic regime, it is natural to expect that ISI arising from the hadronic self-interactions of dark pions can substantially reshape the annihilation cross section. Indeed, as shown in Sec.~\ref{subsec:mf-variation}, the self-scattering amplitude of the dark pions develops resonance poles whose locations depend on the coupling $m_\pi/f_\pi$. By analogy with the Sommerfeld effect, the annihilation cross section should inherit these features via the elastic-channel dressing and thus receive resonant contributions. In what follows, we investigate this expectation quantitatively.

We now compute the ISI contribution to the annihilation amplitude of the dark-pion WIMP. The ingredients required for the analysis are already in place: in the coupled-channel chiral unitary method, we take the interaction kernel $V_\ell(s)$ in each channel to be the LO amplitude, $T^{\rm LO}_{\ell,\pi \pi}(s)$ and $T^{\rm LO}_{\ell,\pi X}(s)$, and we use the standard loop functions for the dark pion $\pi$ and for the SM species $X$. However, since our present interest is the ISI sourced by rescattering in the initial $\pi\pi$ state, we neglect rescattering effects originating from the $X$ sector. This is the same approximation commonly employed in treatments of the Sommerfeld effect and in hadronic ISI. Under these assumptions, the annihilation amplitude $T_{\ell,\pi X}(s)$ takes the form 
\begin{align}
    \label{eq:WIMP ISI}
    T_{\ell,\pi X}(s) \simeq \frac{T^{\rm LO}_{\ell,\pi X}(s)}{1+T^{\rm LO}_{\ell, \pi\pi}(s)G_\pi(s)},
\end{align}
where $G_\pi(s)$ is the loop function in Eq.~(\ref{eq:loop function in DR}) with natural-value subtraction Eq.~(\ref{eq:natural value}) in which the cutoff is take to be the chiral scale $\Lambda_\chi = 4\pi f_\pi$. Since the numerator is the LO annihilation amplitude, the factor multiplying the denominator encodes the ISI effect. Indeed, in the limit where self-scattering is switched off, Eq.~(\ref{eq:WIMP ISI}) reduces to the LO expression. This ISI factor is physically sensible: in the self-scattering sector, neglecting interactions with $X$, the chiral unitary method reads $T_{\ell,\pi \pi}(s)=\big[1+T^{\rm LO}_{\ell, \pi\pi}(s)G_\pi(s)\big]^{-1}T^{\rm LO}_{\ell, \pi\pi}(s)$. Thus the elastic amplitude is precisely the LO self-scattering amplitude $T_{\ell,\pi \pi}(s)$ multiplied by the same ISI dressing factor $\Omega_{\rm ISI}(s) = \big[1+T^{\rm LO}_{\ell, \pi\pi}(s)G_\pi(s)\big]^{-1}$. It follows that the factor introduced earlier is nothing but the elastic rescattering dress of the initial $\pi \pi$ state, and, by the same logic, the annihilation amplitude Eq.~(\ref{eq:WIMP ISI}) takes the form “(ISI dressing) × (tree-level annihilation kernel),” i.e. the short-distance annihilation is modified through the universal elastic $\pi\pi$-rescattering factor. This form, of course, automatically enforces the phase of the elastic self-scattering—exactly as required by Watson’s theorem—so that the annihilation amplitude carries the same phase.

We now make a brief quantitative check of how including ISI, $\Omega_{\rm ISI}(s)$, in the annihilation amplitude feeds into phenomenological predictions for a dark-pion WIMP. As an illustrative example, consider a setup in which the annihilation channel $\pi\pi \to X X$ is $s$-wave dominated. Under this assumption, the annihilation cross section including ISI, weighted by the Maxwell–Boltzmann distribution —an effective quantity that accounts for the velocity dispersion of nonrelativistically moving dark-pion WIMPs—factorizes into the product of the annihilation cross section without ISI and the Maxwell–Boltzmann–weighted ISI factor $\braket{\Omega_{\rm ISI}}(x)$. Since the annihilation cross section is a key ingredient in DM phenomenology over a wide range of velocity dispersions, the corresponding weighted ISI factor $\braket{\Omega_{\rm ISI}}(x)$ can likewise play an important role across different dynamical regimes. For example, the typical relative velocity of DM spans several orders of magnitude across cosmological epochs, from $v_{\rm rel}\sim 10^{-16}$ at the epoch of recombination to $v_{\rm rel}\sim 10^{-3}$ in present-day dark-matter halos. When translated into the Maxwell–Boltzmann variable $x$, these correspond to $x\sim 10^{16}$ and $x\sim 10^{6}$, respectively, and both regimes can impose important constraints on DM models. In the former case, observations of the CMB power spectrum require that entropy injection into the electromagnetic plasma from dark-matter annihilation during recombination be sufficiently suppressed, while in the latter case indirect-detection searches constrain the annihilation cross section not to exceed the observationally allowed limits\,\cite{Kawasaki:2021etm}.

\begin{figure}[t]
\centering
\includegraphics[width=0.7\linewidth]{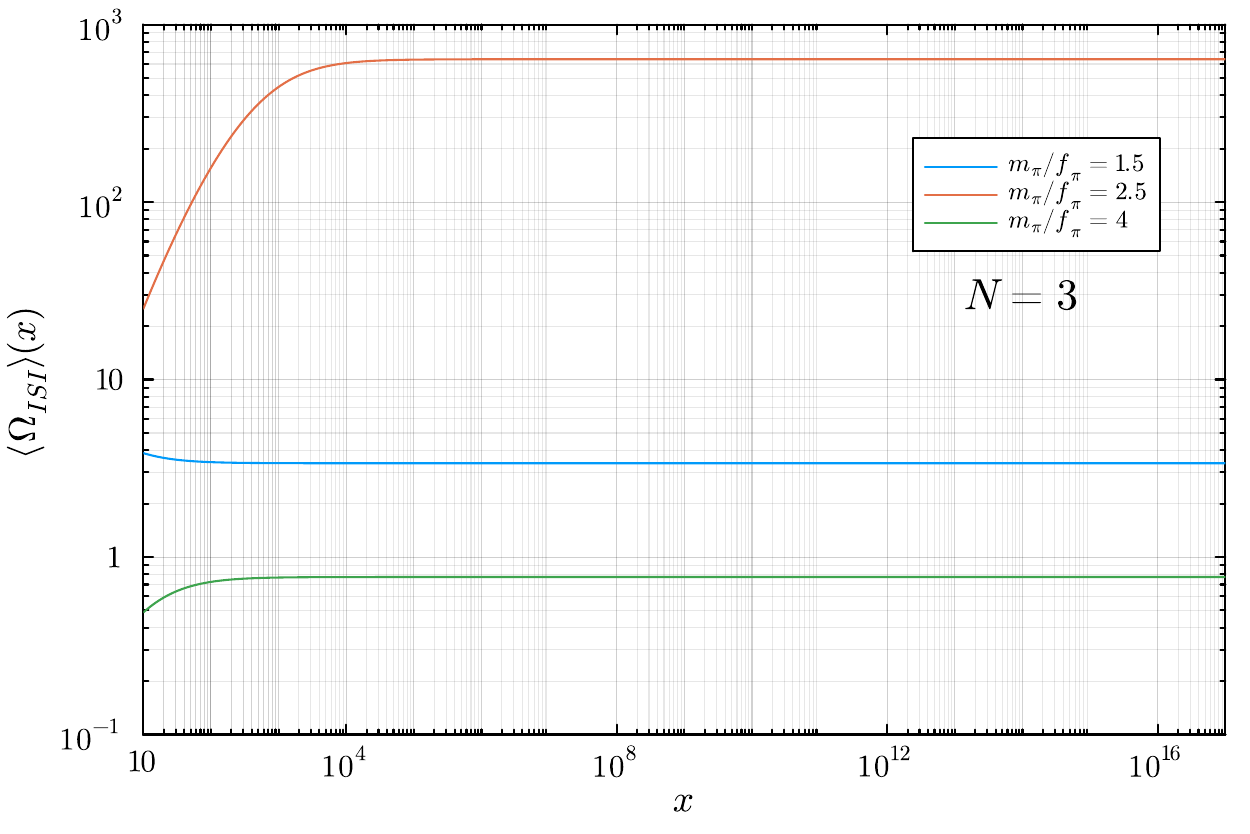}
    \caption{\small \sl 
    Maxwell–Boltzmann–averaged ISI factor $\braket{\Omega_{\rm ISI}}(x)$ as a function of the Maxwell–Boltzmann parameter $x$, which is related to the typical DM relative velocity $v_{\rm rel}$ through $\sqrt{\langle v_{\rm rel}^2 \rangle} = \sqrt{6/x}$, for the symmetry-breaking pattern $SU(N)\times SU(N)/SU(N)$ with $N=3$. The curves correspond to the coupling ratios $m_\pi/f_\pi = 1.5, 2.5,$ and $4.0$.
    }
    \label{fig: ISI}
\end{figure}

With these considerations in mind, we show in the left panel of Fig.~\ref{fig: ISI} the Boltzmann-weighted ISI factor $\braket{\Omega_{\rm ISI}}(x)$ as a function of $x$ for the symmetry-breaking pattern $SU(N)\times SU(N)/SU(N)$ with $N=3$, evaluated at the coupling ratios $m_\pi/f_\pi = 1.5, 2.5,$ and $4$. The factor $\braket{\Omega_{\rm ISI}}(x)$, which encodes the correction to the tree-level annihilation process, saturates at sufficiently large values of $x$, corresponding to the low-velocity regime. In this limit, it approaches $\braket{\Omega_{\rm ISI}}\simeq 3.3$ for $m_\pi/f_\pi=1.5$, $\braket{\Omega_{\rm ISI}}\simeq 600$ for $m_\pi/f_\pi=2.5$, and $\braket{\Omega_{\rm ISI}}\simeq 0.8$ for $m_\pi/f_\pi=4.0$. This demonstrates that the annihilation cross section can receive nontrivial ISI-induced modifications, which can play an important role in shaping low-velocity DM phenomenology, in close analogy with Sommerfeld enhancement from light mediators. As discussed in the previous section, the dependence of the ISI effect on $m_\pi/f_\pi$ can be qualitatively understood in terms of a dynamical resonance present in the self-scattering channel. In the weak-coupling regime, the resonance has a large width and significantly distorts scattering even near threshold, leading to an $\mathcal{O}(1)$ correction. As the coupling increases, the resonance moves closer to threshold, and the scattering exhibits a characteristic resonant behavior. For even stronger coupling, the resonance is pushed below threshold due to the large binding energy, leaving only its tail to influence the scattering amplitude.

The impact of ISI on the annihilation cross section can further lead to nontrivial modifications of the freeze-out process, and hence of the resulting relic abundance. Indeed, as shown in Fig.~\ref{fig: ISI}, for coupling ratios near the resonant region, $m_\pi/f_\pi=2.5$, the ISI factor reaches $\braket{\Omega_{\rm ISI}}(x)\sim\mathcal{O}(100)$ at the typical velocity dispersion relevant for freeze-out, corresponding to $x\sim 20$–$30$. Such a large enhancement is expected to significantly alter the relic abundance determined by the freeze-out mechanism. To examine this effect more concretely, we consider the approximate formula for the relic abundance of WIMP DM in Eq.~(\ref{eq:relic WIMP}). The relic abundance is governed by the late-time depletion that persists after DM decouples from the SM thermal bath, which is encoded in the integral $J(x_f)$. Assuming $s$-wave–dominated annihilation, the tree-level annihilation cross section $(\sigma^{\rm ann}v_{\rm rel})^{\rm ISI}$ that reproduces the observed relic abundance, $\Omega_{\rm WIMP} h^2 \simeq 0.12$, when ISI effects are included in the freeze-out process, is given by 
\begin{align}
    (\sigma^{\rm ann}v_{\rm rel})^{\rm ISI} = \mathcal{E}^{-1}(\sigma^{\rm ann}v_{\rm rel})_{\rm WIMP}, \quad \mathcal{E} = x_f\int^\infty_{x_f} \frac{dx}{x^2} \braket{\Omega_{\rm ISI}}(x),
\end{align}
where $(\sigma^{\rm ann} v_{\rm rel})_{\rm WIMP}$ denotes the canonical WIMP annihilation cross section appearing in Eq.~(\ref{eq:relic WIMP}), and $x_f = m_\pi/T_f$ is the inverse freeze-out temperature.

\begin{figure}[t]
\centering
\includegraphics[width=0.7\linewidth]{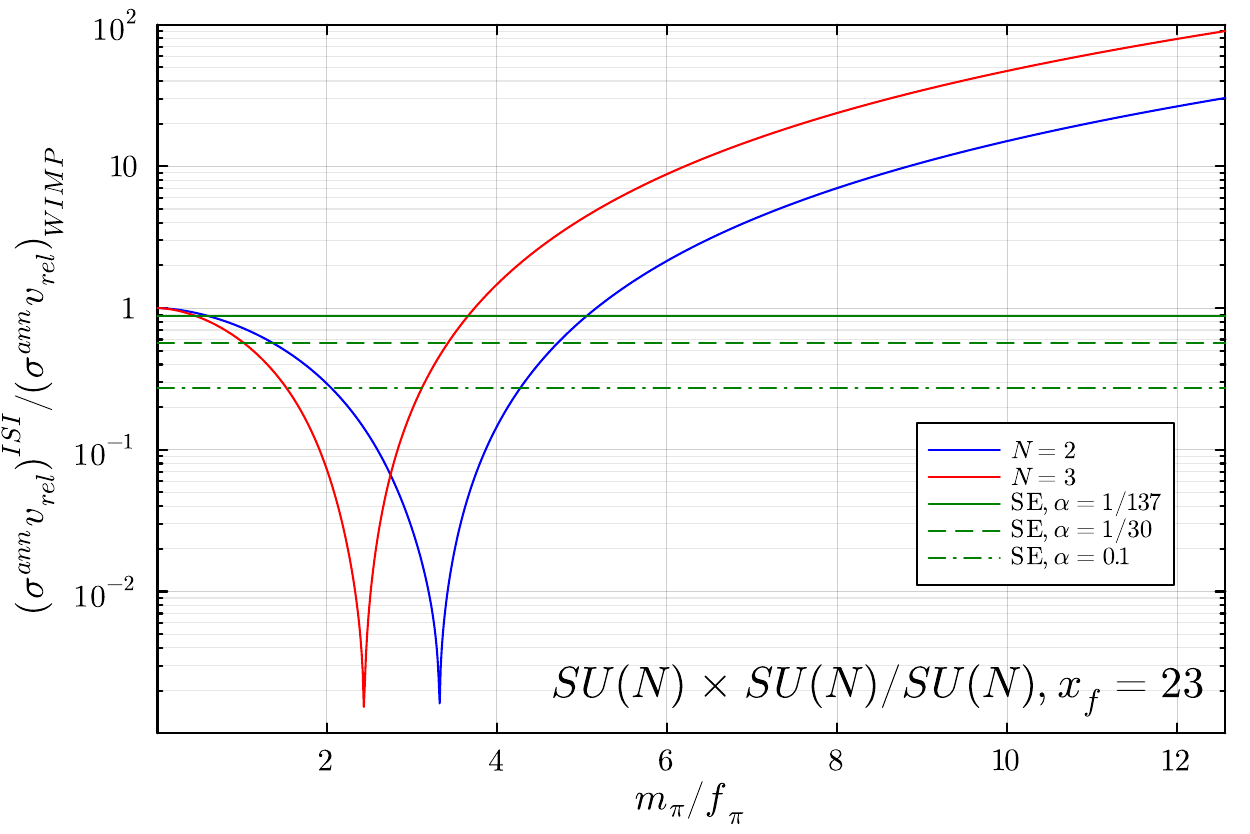}
    \caption{\small \sl 
    Ratio of the annihilation cross section required to reproduce the observed relic abundance when ISI is included to the canonical WIMP annihilation cross section, $(\sigma^{\rm ann} v_{\rm rel})^{\rm ISI}/(\sigma^{\rm ann} v_{\rm rel}){\rm WIMP}$, as a function of the coupling ratio $m\pi/f_\pi$, evaluated at the freeze-out parameter $x_f=23$ and for the symmetry pattern $SU(N)\times SU(N)/SU(N)$. The red (blue) solid curve corresponds to $N=2$ ($N=3$). For comparison, we also show scenarios in which the hadronic ISI between dark pions is switched off and a Coulomb-like interaction is present, characterized by the Sommerfeld factor in Eq.~(\ref{eq:SE Coulomb}); the corresponding deviations from the canonical WIMP cross section are shown by the green solid ($\alpha=1/137$), dashed ($\alpha=1/30$), and dash–dotted ($\alpha=0.1$) curves.
    }
    \label{fig: WIMP mass}
\end{figure}

In Fig.~\ref{fig: WIMP mass}, we plot the ratio $(\sigma^{\rm ann}v_{\rm rel})^{\rm ISI}/(\sigma^{\rm ann}v_{\rm rel})_{\rm WIMP}$ as a function of the coupling constant $m_\pi/f_\pi$, evaluated at freeze-out parameter $x_f=23$ and symmetry pattern $SU(N)\times SU(N)/SU(N)$. The red solid curve shows $N=2$, and the blue solid curve shows $N=3$. For comparison, we also consider a scenario in which the hadronic ISI between dark pions is switched off and a Coulomb-like interaction is present, characterized by the Sommerfeld factor in Eq.~(\ref{eq:SE Coulomb}); the corresponding mass shifts are shown as green solid ($\alpha=1/137$), green dashed ($\alpha=1/30$), and green dash–dot ($\alpha=0.1$) curves. Compared to the deviation induced by the Sommerfeld effect from the canonical WIMP annihilation cross section, we find that hadronic ISI can lead to a larger modification. This is because the typical DM velocity around freeze-out is $v_{\rm rel}\sim 0.5$, i.e. not deeply nonrelativistic, rendering the Sommerfeld effect relatively inefficient. In contrast, in hadronic systems a resonance can form at energies slightly above threshold, enhancing the annihilation rate in the kinematic regime relevant for freeze-out.\footnote{For a long-range interaction described by a Yukawa potential, one can choose parameters such that a shallow bound state appears near threshold. In this case, Sommerfeld enhancement can also have a significant impact on the freeze-out process. This effect is conceptually equivalent to the enhancement discussed in the main text, in the sense that a near-threshold composite state strongly distorts the scattering amplitude.} We emphasize, however, that the present analysis assumes that DM remains in kinetic equilibrium with the thermal bath until chemical freeze-out. This assumption may be violated in parameter regions where a resonance is present, because the coupling constant controlling the tree-level annihilation cross section must be taken to be small in order to reproduce the observed relic abundance. This simultaneously implies a suppression of elastic scattering processes responsible for maintaining kinetic equilibrium between DM and the SM bath. In such a situation, the relic abundance can no longer be reliably described by the Boltzmann equation reduced to the number density alone, as in Eq.~(\ref{eq:Boltzmann WIMP}). Therefore, our purpose here is simply to show that ISI can affect the freeze-out of dark-pion WIMPs, while a concrete model study would require taking early kinetic decoupling\,\cite{Binder:2017rgn} into account.

We briefly comment on how the present treatment of ISI via $\pi\pi$ scattering differs from the standard treatment of the Sommerfeld effect. The Sommerfeld enhancement arises from long-range forces mediated by light carriers, where ladder exchanges of the mediator dominate; accordingly, an $s$-channel bubble resummation that unitarizes a tree-level kernel, which is actually performed in the chiral unitary method, does not by itself capture the characteristic velocity dependence or the Jost/Sommerfeld factors including shallow bound/virtual states. By contrast, in $\pi\pi$-like systems that are right-hand-cut dominated with negligible long-range components, the chiral unitary method with LO kernels provides a good first approximation, preserving the elastic phase and analytic structure while reproducing the resonance onset. Therefore, when Sommerfeld effects are operative, one should explicitly reconstruct the ladder exchange processes by treating the long-range interaction as a nonrelativistic potential-scattering problem and incorporating the corresponding Jost function. A fully consistent assessment of ISI in systems where long-range and hadronic interactions coexist remains a challenging and compelling problem for future work.

\subsection{Subtraction constant and compositeness}
\label{sec:compositeness}

In the analysis so far, we have consistently used the natural-value estimation for the subtraction constant in Eq.~(\ref{eq:natural value}) that enters the loop function $G(s)$. As we have repeatedly noted, however, this quantity is a phenomenological model parameter, so there is no requirement that it actually take the natural value. That said, when the phenomenologically determined subtraction constant deviates substantially from the natural estimate, the deviation itself is highly informative. We now discuss this point in the context of QCD.

Let us consider the scattering of $\pi\pi$ in the isospin $I=1$ channel in QCD, taking the interaction kernel $V_1^1(s)$ to be the tree-level amplitude in Eq.~(\ref{eq:LO amplitudes}) and unitarizing it with the chiral unitary method in Eq.~(\ref{eq:chiral unitary}) as 
\begin{align}
    T_1^1(s) = \left(\frac{6f_\pi^2}{s-4m_\pi^2} + G(s)\right)^{-1},
\end{align}
where the subtraction constant in the loop function $G(s)$ is left as a free parameter. Since numerous experiments establish the existence of the $\rho$ meson, $\sqrt{s_\rho} \simeq 0.775 - i\,0.075\,{\rm GeV}$, in the $I=1$ channel\,\cite{ParticleDataGroup:2024cfk}, we shall determine the subtraction constant phenomenologically by imposing that the above amplitude—analytically continued to the unphysical sheet—possesses this resonance as a pole. This yields
\begin{align}
    \sqrt{s_\rho} \simeq 0.779-i\,0.073\,{\rm GeV}, \quad a(\mu) \simeq -13.8.
\end{align}
If one attempts to reproduce this phenomenologically determined subtraction constant using the natural-value estimation, then, taking the renormalization scale to be $\mu=1\,{\rm GeV}$, one is forced to introduce a cutoff around $\Lambda \simeq 500\,{\rm GeV}$. Imposing such a large cutoff on ChPT is is clearly unphysical, however, because we know that the effective theory of hadrons is UV-completed by QCD around $\Lambda \simeq 2\,{\rm GeV}$.\,\footnote{When the subtraction constant is fixed according to its natural-value estimate, using the $\rho$-meson mass, $\Lambda \simeq 770,{\rm MeV}$, as the cutoff, the resonance pole of the unitarized amplitude $T^{1\mathrm{II}}_1(s)$ is located at $\sqrt{s_\rho} \simeq 0.992 - i1.73\,{\rm GeV}$. Since the corresponding width is very large and the pole lies far from the real axis, it cannot be identified with the physical $\rho$ meson.}

The unnatural cutoff or, equivalently, the large phenomenological subtraction constant, can be traced to the microscopic origin of the $\rho$ meson\,\cite{Oller:1998zr,Oller:2000ma,Oller:2019opk}. If one augments the chiral Lagrangian in Eq.~(\ref{eq:chiral lagrangian}) with an explicit vector field of $\rho$ and computes the tree-level scattering amplitude in the $I=1$ and $p$-wave $\pi\pi$ channel, neglecting $t$ and $u$-channel exchange for clarity, it takes the form\,\cite{Ecker:1988te,Harada:2003jx}
\begin{align}
    T^1_1(s) \simeq \frac{s-4m_\pi^2}{6f_\pi^2}\left[1+g_v^2\frac{s}{m_\rho^2-s}\right].
\end{align}
Here, the $s$-channel exchange of $\rho$ contributes to the second term and the parameter $g_v$ is related to the $\rho \pi \pi$ coupling $g_{\rho \pi \pi}$ as $g_{\rho \pi \pi}=g_v m_\rho/\sqrt{2}f_\pi$, in which the limit $g_v^2 \to 1$ reproduces the so-called KSRF relation\,\cite{Kawarabayashi:1966kd,Riazuddin:1966sw}. Examining the structure of this amplitude for the application of the chiral unitary method, we find that, compared to the case without $\rho$ exchange, it contains an extra zero, or in terms of the inverse amplitude, a new CDD pole: beyond the $p$-wave zero, $s-4m_\pi^2$, already present in the pure $\pi\pi$ amplitude Eq.~(\ref{eq:LO amplitudes}), the explicit $\rho$-exchange term introduces another CDD pole. Including this CDD pole contribution in the interaction kernel $V_1^1(s)^{-1}$, which reduces to a constant in the limit $g_v^2 \to 1$, we obtain
\begin{align}
    T_1^1(s) = \left(\frac{6f_\pi^2}{s-4m_\pi^2}-\frac{6f_\pi^2}{m_\rho^2} + G(s)\right)^{-1}.
\end{align}
The amplitude, when analytically continued to the unphysical sheet, reproduces the $\rho$ pole $\sqrt{s_\rho} \simeq 0.769-i\,0.069\,{\rm GeV} $ with the phenomenologically required subtraction constant $a(1\,{\rm GeV}) \simeq -0.88$, estimated by a natural cutoff of ChPT, $\Lambda_\chi \simeq 770\,{\rm MeV}$. This is due to the fact that constant terms can be consistently shifted between the kernel $V_1^1(s)^{-1}$ and the subtraction constant $a(\mu)$ in the loop function $G(s)$; in the kernel constructed from the ChPT without an explicit $\rho$ field, the same effect is mimicked by driving the subtraction constant to an unnaturally large value. Thus, the deviation of the subtraction constant from the natural-value estimate in Eq.~(\ref{eq:natural value}) is best interpreted as evidence for a missing CDD pole contribution in the kernel $V_1^1(s)^{-1}$—that is, for physics not generated by the $\pi\pi$ rescattering effect alone. This aligns with the standard QCD picture in which the $\rho$ is predominantly described by the $q \bar{q}$ state, whereas the broad $I=0$ scalar $\sigma/f_0(500)$ shows strong indications of a dynamically generated $\pi \pi$ resonance or tetraquark components\,\cite{Pelaez:2015qba,Pelaez:2003dy}. Analogous analyses exist in the meson–baryon sector\,\cite{Hyodo:2008xr,Hyodo:2011ur,Jido:2003cb}, where resonances are classified according to whether they originate from CDD poles (elementary states) or are dynamically generated through nonperturbative resummation of scattering amplitudes.

The implication for dark-pion DM phenomenology is immediate. From a bottom-up perspective, the subtraction constant serves as a model parameter; however, a sizable departure from its natural estimate should be interpreted as evidence for additional elementary states in the dark sector that are not captured by a ChPT constructed solely from dark pions. Neglecting such states may distort annihilation rates and, consequently, relic-abundance predictions, indicating that an LO-only, pion-only framework is inadequate in that channel and should be supplemented by explicit resonance (CDD) degrees of freedom, or by an equivalent representation in the kernel.

It should be emphasized that the discussion in this section does not presuppose the microscopic origin of any particular resonance. The purpose of considering $\pi\pi$ scattering in QCD here is to provide a well-understood illustrative example showing how a deviation of the subtraction constant, as inferred from experiments or phenomenological fits, from its natural-value estimate can be interpreted as a manifestation of a missing CDD pole. Indeed, regarding the origin of the $\sigma/f_0(500)$ resonance, in addition to the interpretation as a dynamically generated state through strong $\pi\pi$ rescattering discussed in this work, there also exist viable approaches in which it is described as a CDD pole, such as in linear sigma models\,\cite{Kondo:2022lgg}. Which description is more appropriate depends on the details of the underlying QCD-like ultraviolet completion. In a dark sector, the microscopic origin of resonances is a priori unknown, and there is no guarantee that the classification realized in QCD—namely, that the $\sigma$ is predominantly dynamical while the $\rho$ is elementary at the pion level—remains valid.

In summary, the key point of this section is that the subtraction constant is not merely a fitting parameter, but rather encodes physical information about which degrees of freedom are explicitly included in a given low-energy effective theory; a substantial deviation from its natural-value estimate should therefore be interpreted as an indication of resonance components not generated by rescattering effects alone, corresponding to missing CDD-pole contributions.

\section{Conclusion}
\label{sec:conclusion}

Dark pions are expected to be the lowest-lying excitations when the dark sector exhibits QCD-like confining dynamics analogous to those of the visible sector, making them compelling DM candidates. Their low-energy interactions are then essentially fixed by the spontaneous symmetry-breaking pattern of the dark sector and are described by ChPT. In practice, however, systematic computations based on this structure are technically demanding, and phenomenological studies in which DM is realized as dark pions are typically carried out at LO in ChPT.

In this work, we revisited this common LO treatment and emphasized that, even in QCD, LO ChPT predictions near threshold—where the expansion is usually regarded as reliable—can differ from high-precision determinations of the isoscalar $s$-wave $\pi\pi$ amplitude by an $\mathcal{O}(1)$ factor. Since this discrepancy is known to originate from the dynamically generated $\sigma$ resonance induced by $\pi\pi$ rescattering, we examined analogous effects in dark-pion systems by employing a nonperturbative resummation of the scattering amplitudes—chiral dynamics—that enforces the correct analytic structure and assessed how the resulting modifications feed into DM phenomenology.

Because unitarization schemes are not unique, we adopted the chiral unitary method, which reproduces empirical $\pi\pi$ scattering with minimal input. This framework required only the LO amplitude and a single model parameter, the subtraction constant. We fixed this constant to its natural estimate by interpreting it as an effective cutoff at the chiral scale $\Lambda_\chi=4\pi f_\pi$, where the dark-pion description ceases to be valid. Consequently, the only free parameters entering the unitarized dark-pion amplitude were the dark-pion mass $m_\pi$ and decay constant $f_\pi$, allowing us to isolate unitarization effects with minimal assumptions. Although quantitative details depend on the cutoff, the qualitative features of the unitarized amplitudes are relatively insensitive to its choice, so the chiral unitary method serves as a good starting point beyond LO.

We first validated the chiral unitary method in the $I=0$ channel of QCD pions and then applied it to dark-pion models, in which the coupling ratio $m_\pi/f_\pi$ is not known a priori. Depending on the ratio $m_\pi/f_\pi$, the amplitudes improved by the chiral unitary method developed resonance poles that are absent in simple perturbative calculations, leading to significant departures from LO results. To clarify the phenomenological implications, we considered two representative realizations of dark-pion DM: SIMP and WIMP. In the SIMP case, the relevant processes were $3\to2$ freeze-out and $2\to2$ self-scattering. Since full unitarization of the $3\to2$ channel is technically challenging, we followed the standard approximate treatment and focused on unitarization effects in the $2\to2$ amplitude. We found that near-threshold resonances can substantially modify the cross section relative to the LO prediction, rendering dark-pion self-interactions highly nontrivial. In the WIMP realization, by contrast, the dark-pion annihilation amplitude exhibited ISI effects analogous to the Sommerfeld enhancement known from light-mediator models. These effects significantly modified the annihilation cross section and thereby altered the phenomenology of dark-pion WIMPs.

Throughout this analysis, the subtraction constant was fixed to its natural estimate. From a model-building perspective, this constant is a phenomenological parameter; however, a substantial deviation from naturalness may be interpreted as evidence for elementary excitations with the same quantum numbers as the relevant channel. This, in turn, suggests that a reliable analysis of dark-pion systems may require going beyond plain ChPT by combining nonperturbative resummation with explicit dark-meson resonance degrees of freedom that cannot be captured in a purely perturbative framework.

\acknowledgments
The author would like to express sincere gratitude to Shigeki Matsumoto for carefully reading the draft and providing valuable comments.


\bibliographystyle{JHEP}
\bibliography{biblio}
\end{document}